\documentclass[10pt,twocolumn,preprintnumbers,amsmath,amssymb,nofootinbib
,superscriptaddress, floatfix]{revtex4-2}

\usepackage{graphicx}
\usepackage{aas_macros}
\usepackage{dcolumn}
\usepackage{bm}
\usepackage[utf8]{inputenc}
\usepackage{color}
\usepackage[colorlinks,linkcolor=blue,citecolor=red,urlcolor=blue ]{hyperref}
\usepackage[dvipsnames]{xcolor}
\usepackage{mathptmx}
\usepackage{xspace}
\usepackage{booktabs}

\newcommand{\emis}{\dot{n}^\gamma}
\newcommand{\emiss}[1]{\langle b\emis_{#1}\rangle}
\newcommand{\gammaray}{$\gamma$-ray }
\newcommand{\gammarays}{$\gamma$-rays }
\newcommand{\sigv}{\langle\sigma v\rangle}
\newcommand{\unitn}{\hat{\textbf{n}}}
\newcommand{\Fermi}{{\sl Fermi}-LAT\xspace}
\newcommand{\wisc}{WI-SC\xspace}

\defcitealias{1312.1729}{SC14}
\defcitealias{1603.04057}{M16}
\defcitealias{1107.1916}{G12}

\begin{document}

\title{Constraints on dark matter and astrophysics from tomographic $\gamma$-ray cross-correlations}

\author{Anya Paopiamsap}
\email{anyabua@gmail.com}
\affiliation{Department of Physics, University of Oxford, Denys Wilkinson Building, Keble Road, Oxford OX1 3RH, United Kingdom}
\author{David Alonso}
\affiliation{Department of Physics, University of Oxford, Denys Wilkinson Building, Keble Road, Oxford OX1 3RH, United Kingdom}
\author{Deaglan J. Bartlett}
\affiliation{Department of Physics, University of Oxford, Denys Wilkinson Building, Keble Road, Oxford OX1 3RH, United Kingdom}
\affiliation{CNRS \& Sorbonne Universit\'e, Institut d’Astrophysique de Paris (IAP), UMR 7095, 98 bis bd Arago, F-75014 Paris, France}
\author{Maciej Bilicki}
\affiliation{Center for Theoretical Physics, Polish Academy of Sciences, al. Lotnik\'{o}w 32/46, 02-668 Warsaw, Poland}
\date{\today}

\begin{abstract}
  We study the cross-correlation between maps of the unresolved $\gamma$-ray background constructed from the 12-year data release of the {\sl Fermi} Large-Area Telescope, and the overdensity of galaxies in the redshift range $z\lesssim0.4$ as measured by the 2MASS Photometric Redshift survey and the WISE-SuperCOSMOS photometric survey. A signal is detected at the $8-10\sigma$ level, which we interpret in terms of both astrophysical $\gamma$-ray sources, and WIMP dark matter decay and annihilation. The sensitivity achieved allows us to characterise the energy and redshift dependence of the signal, and we show that the latter is incompatible with a pure dark matter origin. We thus use our measurement to place an upper bound on the WIMP decay rate and the annihilation cross-section, finding constraints that are competitive with those found in other analyses. Our analysis is based on the extraction of clean model-independent observables that can then be used to constrain arbitrary astrophysical and particle physics models. In this sense we produce measurements of the $\gamma$-ray emissivity as a function of redshift and rest-frame energy $\epsilon$, and of a quantity $F(\epsilon)$ encapsulating all WIMP parameters relevant for dark matter decay or annihilation. We make these measurements, together with a full account of their statistical uncertainties, publicly available.
\end{abstract}

\maketitle

\section{Introduction}
  The unresolved \gammaray background (UGRB) is the collective emission observed after subtracting the diffuse Galactic contribution and detected extragalactic sources. It originates from unresolved astrophysical sources, including star-forming galaxies \cite{2109.07598,2106.07308}, misaligned active galactic nuclei (AGNs) \citep{1304.0908}, blazars \citep{1912.01622,2201.02634}, millisecond pulsars \citep{1406.2706}, and the interaction of cosmic rays with the extragalactic background light \citep{0704.2463} (see \cite{1502.02866} for a review). More interestingly for fundamental physics, the UGRB may also be used for the indirect detection of particle dark matter.

  Weakly-interacting massive particles (WIMPs) are one of the most promising candidates to make up a substantial fraction of the dark matter (DM) needed to explain cosmological observations \citep{1985NuPhB.253..375S,hep-ph/0404175,1807.06209,2007.15632,2105.13549}. With masses in the range of $m_\chi\simeq1\,{\rm GeV}-1{\rm TeV}$\footnote{We use natural units with $c=1$ throughout.}, if produced thermally in the early Universe and then decoupled from the plasma, their relic density would be naturally of the order of the measured global dark matter abundance \citep{1977PhRvL..39..165L,1978ApJ...223.1015G}. In most scenarios, decay or annihilation of WIMPs into Standard Model particles leads to the emissions of \gammaray photons. For this reason, data from \gammaray observatories, such as the \Fermi telescope, has been widely used to place constraints on WIMPs as a dark matter candidate \citep{1605.02016,2208.00145}.

  Both the extragalactic astrophysical sources of the UGRB listed above, as well as any potential contribution from DM processes, would trace the same large-scale structures, and hence cause anisotropies in this background. These anisotropies have been detected and studied via measurements of the UGRB power spectrum \citep{1202.2856,1410.3696,1608.07289,1812.02079}, revealing important information about its energy spectrum and potential composition. However, significant additional information can be obtained from the cross-correlations of the UGRB with other tracers of the same large-scale structure. Firstly, cross-correlations with high signal-to-noise tracers of structure are able to tease out low-significance signatures in the data \citep[e.g.][]{2002.02888,2005.00244}. Secondly, the cross-correlation with tracers at different redshifts allows us to reconstruct the joint dependence of the UGRB on redshift, energy, and physical scales, which is vital to separate the astrophysical sources of the signal from its potential DM contributions.

  In this sense, various probes of structure offer different advantages. The spatial distribution of galaxies is generally the highest signal-to-noise tracer of the matter fluctuations and, if redshift information is available, allows for an accurate tomographic reconstruction of the signal \citep{1407.8502,1506.01030,1503.05922,1511.07092,1709.01940,1808.09225,2205.12916}. The main difficulty, however, is ensuring an accurate modelling of the relationship between galaxy and matter overdensities, particularly on small scales \cite{astro-ph/0207664,1707.03397,2211.01744}. Cosmic shear, caused by the weak gravitational lensing of background galaxies, is a direct tracer of the matter fluctuations, and is therefore immune to this challenge. However, shape measurement noise significantly reduces the sensitivity of this tracer, even for the densest galaxy samples and, as a cumulative effect along the line of sight, separating the \gammaray signal into its contributions at different redshifts is less straightforward \citep{1212.5018,1404.5503,1607.02187,1611.03554,1802.10257,1907.13484}. Another promising cross-correlation tracer is the positions of galaxy clusters. As the most massive objects in the Universe, clusters of galaxies contain significant amounts of dark matter, and hence are promising environments to detect the associated \gammaray signal \citep{1612.05788,1805.08139,1907.05264,2303.16930}. Cross-correlation studies with other less standard tracers have been carried out in the literature, including Cosmic Microwave Background anisotropies (CMB) \citep{2005.03833}, lensing of the CMB \citep{1410.4997}, the Cosmic Infrared Background \citep{1608.04351}, the late-time 21cm signal \citep{1911.04989}, the thermal Sunyaev-Zel'dovich effect \citep{1911.11841}, and high-energy neutrino events \citep{2304.10934}. Finally, some of the tightest constraints on WIMPs have been obtained from the study of the UGRB around local structures \citep{1312.7609,1503.02641,1611.03184,2109.11291}.

  In this paper, we study the cross-correlation of maps of the UGRB from 12 years of \Fermi data with tomographic maps of the galaxy overdensity constructed from the 2MASS Photometric Redshift survey (2MPZ) and the WISE-SuperCOSMOS photometric survey (\wisc), covering the redshift range $z\lesssim0.4$. We interpret the associated signal in terms of both astrophysical \gammaray sources and DM processes. The main improvements from our analysis with respect to previous similar works (e.g. \cite{1506.01030,1709.01940}) is the use of newer, more sensitive \Fermi data (12-year dataset as opposed to the 8-year release used in previous works, together with its updated point-source mask), with correspondingly enhanced constraints on DM decay and annihilation, and the adoption of an agnostic modelling approach. This allows us to compress our data into the measurement of a few model-independent quantities\footnote{We make these measurements and their statistical uncertainties publicly available in \url{https://github.com/anyabua/FermiX}.}. As we show, this has key advantages, enabling an easy interpretation of our measurements in the context of arbitrary DM particle interactions, as well as providing robust methods to identify the presence of astrophysical contributions in the data.

  This paper is structured as follows. Section \ref{sec:th} presents the theoretical background used to model the UGRB and its cross-correlation with galaxies. The datasets used in our analysis are described in Section \ref{sec:data}. The data analysis methodology we employ is presented in Section \ref{sec:meth}. Section \ref{sec:res} then presents the results of our analysis, and the corresponding constraints on DM interactions and \gammaray astrophysics. We summarise these and conclude in Section \ref{sec:conc}.

\section{Theoretical model}\label{sec:th}
  \subsection{Power spectra and the halo model}\label{ssec:th.cls}
    Our two observables, the angular galaxy overdensity $\delta_g$ (see Section \ref{ssec:th.gal}), and the $\gamma$-ray intensity $I_\gamma$ (see Section \ref{ssec:th.gamma}), are sky projections of 3-dimensional fields, and can be written in general as
    \begin{equation}
      u(\unitn) = \int d\chi q_{u}(\chi)U(\chi\unitn,z),
    \end{equation}
    where $U$ is the 3D counterpart of the projected field $u$, $\chi$ is the radial comoving distance, and $q_u(\chi)$ is the radial kernel defining the projection. The angular power spectrum of two such fields, $u$ and $v$, can be calculated as
    \begin{equation}
      C_\ell^{uv}=\int\frac{d\chi}{\chi^2}q_u(\chi)q_v(\chi)\,P_{UV}\left(k_\ell,z\right),
    \end{equation}
    where $P_{UV}(k,z)$ is the power spectrum of the associated 3D fields, and $k_\ell\equiv(\ell+1/2)/\chi$. The equation above assumes the applicability of Limber's approximation \cite{1953ApJ...117..134L,1992ApJ...388..272K}, which holds when the radial kernels of the fields involved are significantly broader than their typical correlation length (this is the case for the observables studied here).

    The 3D power spectra can be described using the halo model \citep{astro-ph/0005010,astro-ph/0001493,astro-ph/0206508}. In this formalism, the power spectrum receives two contributions: the ``two-halo'' term, corresponding to the contribution from pairs of matter elements in different haloes, and the ``one-halo'' term, corresponding to matter elements belonging to the same halo:
    \begin{equation}
      P_{UV}(k)=P_{UV}^{2h}(k)+P_{UV}^{1h}(k),
    \end{equation}
    where we have omitted the redshift dependence for brevity. These two contributions are given by
    \begin{align}\label{eq:p1h}
      &P_{UV}^{1h}(k)=\int_{M_{\rm min}} dM\,n(M)\,\langle U(k|M)\,V(k|M)\rangle,\\\label{eq:p2h}
      &P_{UV}^{2h}(k)=\langle bU\rangle\,\langle bV\rangle\,P_L(k),\\\label{eq:bU}
      &\langle bU\rangle\equiv \int_{M_{\rm min}} dM\,n(M)\,b_h(M)\,\langle U(k|M)\rangle.
    \end{align}
    where $n(M)$ is the halo mass function, $b_h(M)$ is the halo bias, $U(k|M)$ is the Fourier transform of the halo profile for field $U$ around a halo of mass $M$, $P_L(k)$ is the linear matter power spectrum, and $\langle\cdots\rangle$ denotes averaging over all possible realisations of haloes of the same mass. The bias-weighted observable $\langle bU\rangle$ will be of particular interest in Section \ref{ssec:th.gamma_astro}.

  \subsection{Galaxy Clustering}\label{ssec:th.gal}
    We cross-correlate the $\gamma$-ray intensity maps with projected maps of the galaxy distribution in a set of tomographic redshift bins. Concretely, we produce maps of the projected galaxy overdensity $\delta_g(\unitn)\equiv n_g(\unitn)/\bar{n}_g-1$, where $n_g$ is the angular number density of galaxies, and $\bar{n}_g$ is its sky average. This quantity is related to the underlying 3D galaxy overdensity $\Delta_g$ via
    \begin{equation}
      \delta_g(\unitn)=\int d\chi\,\frac{dz}{d\chi}\,p_g(z)\,\Delta_g(\chi\unitn),
    \end{equation}
    where $p_g(z)$ is the sample's redshift distribution, and $dz/d\chi=H(z)$ is the expansion rate at redshift $z$.

    For simplicity, we assume that galaxies are a biased tracer of the underlying dark-matter overdensity
    \begin{equation}
      \Delta_g({\bf x})=b_g\,\delta({\bf x}),
    \end{equation}
    where $b_g$ is the galaxy bias. We also assume that dark matter is distributed in haloes according to a truncated Navarro-Frenk-White (NFW) profile \citep{astro-ph/9508025} (see Appendix \ref{app:halo.nfw} for details).

    Thus, within this simple model, the radial kernel and halo profile associated with galaxy clustering are:
    \begin{align}\label{eq:kernel_gal}
      q_g(\chi)=b_g\frac{dz}{d\chi}p_g(z),\hspace{12pt}
      U_g(r)=\frac{\rho_{\rm NFW}(r)}{\bar{\rho}_{M,0}},
    \end{align}
    where $\bar{\rho}_{M,0}$ is the mean matter density today. To test that our final results are not sensitive to the details of the galaxy-halo relation, we repeat parts of our analysis for a more sophisticated model, using the so-called halo occupation distribution approach (HOD). The model is described in more detail in Appendix \ref{app:halo.hod}.

  \subsection{Gamma ray intensity}\label{ssec:th.gamma}
    \gammaray observations can be used to construct maps of the UGRB intensity, defined as the number $N_o$ of observed photons detected per unit time $t_o$, detector area $A_o$, photon energy $\epsilon_o$, and solid angle $d\Omega_o$ along direction $\unitn$:
    \begin{equation}
      I(\unitn,\epsilon_o)=\frac{dN_o}{dt_odA_od\epsilon_od\Omega_o},
    \end{equation}
    where the subscript $_o$ denotes quantities measured in the observer's frame. The contribution to the total observed intensity from sources in a distance interval $dl_e$ along the line of sight is given by
    \begin{equation}
      dI(\unitn)=\frac{dN_e}{dt_ed\epsilon_edV_ed\Omega_e}\frac{dN_o}{dN_e}\frac{dt_ed\epsilon_e}{dt_od\epsilon_o}\frac{dA_ed\Omega_e}{dA_od\Omega_o}dl_e,
    \end{equation}
    where $_e$ denotes quantities in the frame of the emitting material, and the transverse volume element is $dV_e\equiv dA_edl_e$. Ignoring all geometric and redshift distortions, we can write:
    \begin{align}
      \frac{d\epsilon_edt_e}{d\epsilon_odt_o}=1,\hspace{6pt}
      \frac{dA_ed\Omega_e}{dA_od\Omega_o}=\frac{1}{(1+z)^2},\hspace{6pt}
      dl_e=\frac{d\chi}{1+z},
    \end{align}
    where $d\chi$ is an interval of comoving distance. The ratio of observed to emitted photons can for now be written as
    \begin{equation}
      \frac{dN_o}{dN_e}=\exp[-\tau(\chi)],
    \end{equation}
    where the optical depth $\tau$ accounts for the absorption of \gammaray photons by the extragalactic background light \citep{0805.1841}.

    For simplicity, in what follows, we will label energies measured in the observer's frame with an uppercase $E$, and rest-frame energies simply with a Greek lowercase $\epsilon$.

    The total intensity can then be calculated by integrating along the line of sight, obtaining:
    \begin{equation}\label{eq:Iint}
      I(\unitn)=\int d\chi\,e^{-\tau}\frac{\emis_\epsilon(\chi\unitn)}{4\pi(1+z)^3},
    \end{equation}
    where $\emis_\epsilon$ is the emissivity: the physical\footnote{As opposed to comoving.} number density of photons emitted per unit time and energy
    \begin{equation}
      \emis_\epsilon \equiv\frac{dN_e}{dt_ed\epsilon_{e} dV_e}.
    \end{equation}
    The form of $\emis_\epsilon$ (e.g. its dependence on cosmological and astrophysical parameters) depends on the process giving rise to $\gamma$-ray emission. In the next two sections we will present models to describe $\gamma$-ray emission from DM decay and annihilation, and from regular astrophysical processes.

  \subsection{Gamma rays from dark matter}\label{ssec:th.gamma_DM}
    \subsubsection{Dark matter annihilation}\label{sssec:th.gamma_DM.ann}
      The probability per unit time for one DM particle to annihilate against another is
      \begin{equation}
        dp_{\rm ann}=n_{\rm DM}\sigma dx=n_{\rm DM}\langle\sigma v\rangle\,dt,
      \end{equation}
      where $n_{\rm DM}$ is the number density of DM particles, and $\langle\sigma v\rangle$ is the annihilation cross section times the relative velocity between particles averaged over the velocity distribution. The total number of annihilations taking place per unit volume and time is therefore
      \begin{equation}
        \frac{dN_{\rm ann}}{dt\,dV}=\frac{n_{\rm DM}}{2}\frac{dp_{\rm ann}}{dt}=\frac{1}{2}\frac{\rho_{\rm DM}^2}{m_{\rm DM}^2}\langle\sigma v\rangle,
      \end{equation}
      where $\rho_{\rm DM}$ is the DM density, and $m_{\rm DM}$ is the DM particle mass. The factor of $1/2$ must be included to avoid double-counting unique pairs of annihilating particles.

      To obtain the emissivity due to annihilation we simply multiply the number density rate of annihilations by the spectrum $dN_\gamma^{\rm ann}/d\epsilon$ (i.e. the number of photons emitted per unit energy) for each annihilation:
      \begin{equation}
        \emis_\epsilon({\bf x})=\frac{\langle\sigma v\rangle}{2m_{\rm DM}^2}\,\frac{dN_\gamma^{\rm ann}}{d\epsilon}\,\rho_{\rm DM}^2({\bf x}).
      \end{equation}
    
    \subsubsection{Dark matter decay}\label{sssec:th.gamma_DM.dec}
      Calculating the emissivity for DM decay is even simpler. The probability per unit time for a DM particle to decay $dp_{\rm dec}/dt$ is simply given by the decay rate $\Gamma$. Thus, the number of decays per unit volume and time is
      \begin{equation}
        \frac{dN_{\rm dec}}{dV\,dt}=\Gamma\,n_{\rm DM}.
      \end{equation}
      To obtain the emissivity we simply multiply this by the decay spectrum $dN_\gamma^{\rm dec}/d\epsilon$:
      \begin{equation}\label{eq:emiss.ann}
        \emis_\epsilon({\bf x})=\frac{\Gamma}{m_{\rm DM}}\,\frac{dN_\gamma^{\rm dec}}{d\epsilon}\,\rho_{\rm DM}({\bf x}).
      \end{equation}

    \subsubsection{\gammaray intensity maps from DM decay and annihilation}\label{sssec:th.gamma_DM.imaps}
      Combining the results for decay and annihilation, the $\gamma$-ray intensity can be written schematically as
      \begin{align}\label{eq:int_dm_gen}
        I_\gamma(\unitn,E)=\int d\chi\,e^{-\tau(\chi)}(1+z)\,C(z)\,F(\epsilon)\,\Delta_\gamma(\chi\unitn),
      \end{align}
      where $\epsilon\equiv E(1+z)$ is the rest-frame energy of the emitted photons. Above, $C(z)$ is a radial kernel involving only cosmological quantities, $F(\epsilon)$ is a function depending only on particle physics properties of DM, and $\Delta_\gamma({\bf x})$ is a three-dimensional field tracing the large-scale structure. The specific form of these quantities depends on the emission process (decay or annihilation):
      \begin{itemize}
        \item {\bf Annihilation:}
        \begin{align}\label{eq:dm_ann_C}
          &C(z)\equiv\frac{\rho_{c,0}^2\Omega_{\rm DM}^2}{8\pi}(1+z)^2,\\
          &F(\epsilon)\equiv\frac{\langle\sigma v\rangle}{m_{\rm DM}^2}\label{eq:Fann}\frac{dN_\gamma^{\rm ann}}{d\epsilon},\\
          &\Delta_\gamma\equiv (1+\delta)^2.
        \end{align}
        \item {\bf Decay:}
        \begin{align}\label{eq:dm_dec_C}
          &C(z)\equiv\frac{\rho_{c,0}\Omega_{\rm DM}}{4\pi(1+z)},\\
          &F(\epsilon)\equiv\frac{\Gamma}{m_{\rm DM}}\frac{dN_\gamma^{\rm dec}}{d\epsilon}\label{eq:Fdec},\\
          &\Delta_\gamma\equiv 1+\delta.
        \end{align}
      \end{itemize}
      Here, $\delta({\bf x})\equiv\rho_{\rm DM}({\bf x})/\bar{\rho}_{\rm DM}-1$ is the DM overdensity, $\rho_{c,0}\equiv 3H_0^2/8\pi G$ is the critical density today, with $H_0$ the Hubble constant, $\Omega_{\rm DM}\equiv\bar{\rho}_{{\rm DM},0}/\rho_{c,0}$ is the fractional dark matter density, and we have used the fact that the DM density evolves as $\bar{\rho}_{\rm DM}(z)=\bar{\rho}_{{\rm DM},0}(1+z)^3$.

      Finally, we will make use of intensity maps integrated over a finite energy bin
      \begin{align}
        \bar{I}_\gamma^i(\unitn)
        &\equiv\frac{1}{E_{i+1}-E_i}\int_{E_i}^{E_{i+1}}dE\,I_\gamma(\unitn,E).
      \end{align}
      These are then related to the quantities defined above via
      \begin{equation}\label{eq:gamma_int_1}
        \bar{I}_\gamma^i(\unitn)=\int d\chi\,e^{-\tau(\chi)}C(z)\bar{F}_i(z)\Delta_\gamma(\chi\unitn),
      \end{equation}
      where
      \begin{equation}
        \bar{F}_i(z)\equiv\frac{1}{E_{i+1}-E_i}\int_{E_i(1+z)}^{E_{i+1}(1+z)} d\epsilon\, F(\epsilon).
      \end{equation}

      With this, the radial kernel associated with the $\gamma$-ray intensity maps is:
      \begin{align}\label{eq:kernel_DM}
        q_\gamma^i(\chi)=e^{-\tau(\chi)}C(z)\bar{F}_i(z).
      \end{align}
      At the redshifts and energies under study, the optical depth $\tau$ can be safely ignored \citep{0805.1841}, and we will do so in what follows.

      The halo profile associated with DM decay is simply the NFW profile, described in Appendix \ref{app:halo.nfw}:
      \begin{equation}
        U_\gamma^{\rm dec}(r)=\frac{\rho_{\rm NFW}(r)}{\bar{\rho}_{M,0}}.
      \end{equation}
      
      The case of annihilation is more complicated. Since annihilation is proportional to the mean of the squared DM density, the signal is highly sensitive to the amount of substructure, i.e. fluctuations around the mean NFW profile for haloes of a given mass, normally manifested in the form of subhaloes. We will present constraints from annihilation assuming three different models to describe the boost factor to the annihilation halo profile associated with substructures. Ordered by the amplitude of the associated boost factor, these are the substructure models of \cite{1312.1729}, \cite{1603.04057}, and \cite{1107.1916} (labelled SC14, M16, and G12 hereon, respectively). Details of these four models are described in Appendix \ref{app:halo.ann}, and follow the description in \cite{1301.5901}. Unless otherwise stated, our fiducial constraints on annihilation will assume the M16 substructure model.
      
      Another important aspect of the model for annihilation is the minimum mass over which haloes contribute effectively to the signal \cite{1301.5901} (see Eqs. \ref{eq:p1h} and \ref{eq:bU}). As in previous works (e.g. \cite{1506.01030}), we will set this to be $M_{\rm min}=10^{-6}M_\odot$, corresponding to a typical WIMP free-streaming mass. We explore the uncertainty associated with this choice in Appendix \ref{app:Mmin}.

    \subsubsection{Modeling $F$}\label{sssec:th.gamma_DM.F}
      $F(\epsilon)$ depends solely on the fundamental properties of the DM particles: mass, cross section/decay rate, and the associated spectra. The specific form of $F$ thus depends on the particle physics model used to describe DM and its interactions. Here we will take an agnostic approach and instead attempt to reconstruct the form of $F$ directly from the data. To do so, we will model $F(\epsilon)$ simply as a step-wise function:
      \begin{equation}\label{eq:fstep}
        F(\epsilon)\equiv\sum_{n=1}^{N_F} F_n\,\Theta(\epsilon_n<\epsilon<\epsilon_{n+1}),
      \end{equation}
      where $\Theta(\epsilon_n<\epsilon<\epsilon_{n+1})$ is a top-hat function in the range $[\epsilon_n,\epsilon_{n+1})$, and the amplitudes $F_n$ are free parameters of the model. Inserting this into Eq. \ref{eq:gamma_int_1}, the model for the intensity map in the $i$-th bin takes a particularly simple form:
      \begin{equation}
        I_\gamma^i(\unitn)=\sum_{n=1}^{N_F} F_n\,\mathcal{I}_\gamma^{i,n}(\unitn),
      \end{equation}
      where
      \begin{equation}
        \mathcal{I}_\gamma^{i,n}(\unitn)\equiv\int d\chi\,e^{-\tau(\chi)}C(z)W_{i,n}(z)\Delta_\gamma(\chi\unitn),
      \end{equation}
      and
      \begin{align}\nonumber
        W_{i,n}(z)
        &\equiv\frac{1}{E_{i+1}-E_i}\int_{E_i(1+z)}^{E_{i+1}(1+z)}d\epsilon\,\Theta(\epsilon_n<\epsilon<\epsilon_{n+1})\\\nonumber
        &=\frac{\text{Min}[E_{i+1}(1+z),\epsilon_{n+1}]-\text{Max}[E_{i}(1+z),\epsilon_{n}]}{E_{i+1}-E_{i}}\\
        &\,\,\,\,\,\,\times\Theta(\epsilon_{n+1}>E_{i}(1+z))\Theta(\epsilon_{n}<E_{i+1}(1+z)).
      \end{align}
      Above, the two last Heavyside functions ensure that the integral is zero when there is no overlap between both energy intervals.

      For simplicity, and since we will only explore relatively low redshifts ($z\lesssim0.4$), and hence the effects of redshifting from the source to the observer frame are mild relative to the energy bin widths, we will use the same bin boundaries used to construct the intensity maps, $\{E_i\}$, as the edges $\{\epsilon_n\}$ used to model $F$ in Eq. \ref{eq:fstep}.

    Finally, note that the step-wise parametrisation of $F(\epsilon)$ (Eq. \ref{eq:fstep}) allows us to write the cross-correlation between the $g$-th galaxy sample and the $i$-th \gammaray intensity map as
    \begin{equation}\label{eq:F_parametrisation}
      C^{g,i}_{\ell} = \sum_{n}F_{n}C^{g,i,n}_{\ell},
    \end{equation}
    where
    \begin{equation}\label{eq:cl_parametrisation}
      C^{g,i,n}_{\ell}\equiv\int\frac{dz}{\chi^2}b_gp_g(z)\,C(z)W_{n,i}(z)\,P_{\delta\Delta_\gamma}(k_\ell,z).
    \end{equation}
    The template power spectra $C^{g,i,n}_\ell$ thus depend solely on cosmological quantities (distances, redshifts, and power spectra of various powers of the matter overdensity), while all the particle-physics properties are compressed into the linear amplitudes $F_n$. Our aim will therefore be to reconstruct $F_n$ from the measured cross-correlations.

    \subsection{Astrophysical \gammaray emission}\label{ssec:th.gamma_astro}
      Since astrophysical \gammaray sources trace the same large-scale structure as the dark matter structures that could contribute to the UGRB through decay or annihilation, their contribution may dominate the cross-correlation with the galaxy overdensity, and is the main contaminant for this type of study. Here we develop a simple and generic scheme, based on the halo model, to interpret our measurements purely in terms of an astrophysical signal. Our treatment follows closely that used by \cite{2206.15394} to constrain the star formation rate density from cross-correlations of the Cosmic Infrared Background.

      Consider Eq. \ref{eq:p2h} in the large-scale limit (i.e. on scales larger than the typical size of a halo). In this regime, we can write the 2-halo contribution to the galaxy-\gammaray cross power spectrum as
      \begin{equation}
        P_{g\gamma}^{2h}(k)=\emiss{\epsilon}\,b_g\,P_{mm}(k),
      \end{equation}
      where $b_g$ is the galaxy bias, and we have defined the \emph{bias-weighted mean $\gamma$-ray emissivity}
      \begin{equation}
        \emiss{\epsilon}\equiv\int dM\,n(M)\,b_h(M)\,\frac{d\dot{N}}{d\epsilon}(M),
      \end{equation}
      with $d\dot{N}/d\epsilon$ the specific luminosity (total number of photons per unit time and energy interval emitted by the halo). Assuming we know the galaxy bias $b_g$, and the matter power spectrum $P_{mm}(k)$, the amplitude of the cross-correlation is therefore sensitive to the mean $\gamma$-ray emissivity (weighted by halo bias).

      The 1-halo term is more difficult to interpret since it depends on the covariance between the galaxy density and the $\gamma$-ray emissivity within haloes. However, following \cite{1709.01940,2206.15394}, we can assume the scale dependence of this term to be effectively flat in harmonic space. This is an acceptable approximation since the \textit{Fermi} point-spread function (PSF) is too broad to resolve dark-matter haloes in detail.

      Using this approximation, and assuming the galaxy samples under study have relatively narrow redshift bins (compared to the typical variation of $\emiss{\epsilon}$), we can write a remarkably simple model for the cross-power spectrum between the $g$-th sample of galaxies and the $i$-th $\gamma$-ray intensity map, at energy $E_i$:
      \begin{equation}\label{eq:model_astro}
        C_\ell^{g,i}=\emiss{E_i(1+z_g)}\,Q_\ell^{g}+C^{g,i}_{\rm 1h},
      \end{equation}
      where $z_g$ is the mean redshift of the sample, $C^{g,i}_{\rm 1h}$ is the 1-halo contribution to this power spectrum, and the template $Q_\ell^g$ is (see Eq. \ref{eq:Iint})
      \begin{equation}\label{eq:Q_astro}
        Q_\ell^g\equiv \int\frac{dz}{\chi^2}\frac{b_gp_g(z)}{4\pi(1+z)^3}P_{mm}(k_\ell,z).
      \end{equation}
      Note that $\emiss{\epsilon}$ in Eq. \ref{eq:model_astro} is evaluated at the rest-frame energy $E(1+z_g)$, and that itself is an intrinsic function of redshift (which we do not state explicitly above for brevity).

      We can thus reconstruct the energy and redshift dependence of the $\gamma$-ray emissivity by fitting the model in Eq. \ref{eq:model_astro} to the measured galaxy-$\gamma$-ray correlations using $\emiss{\epsilon}$ and $C^{g,E}_{\rm 1h}$ as free parameters.

\section{Data}\label{sec:data}
  \subsection{The \Fermi 12-year data}\label{ssec:data.fermi}
    In this work we use 12 years of \gammaray observations from the \textit{Fermi} Large Area Telescope (\Fermi), which we process using the {\it Fermi} Tools and {\tt FermiPY} \citep{1707.09551}\footnote{\url{https://github.com/fermiPy/fermipy}.}, following a similar procedure to \citet{1812.02079}. We reject the lowest quartile of photons according to their PSF (PSF0) and divide the remaining photons into 100 logarithmically spaced energy bins in the range $[0.5248,1000] \,\,{\rm GeV}$. To project and bin these observations spatially onto a sky map, we use the {\tt healpy} package \cite{2019JOSS....4.1298Z} and the {\tt HEALPIX} pixelation scheme. More information on {\tt HEALPIX} can be found in \cite{astro-ph/0409513}. The angular resolution of the data (parametrised by the {\tt HEALPIX} resolution parameter {\tt nside}), is {\tt nside} = 1024 which corresponds to a pixel size of $\sim 3.4 $ arcminutes.

    \begin{figure}[t]
        \centering
        \includegraphics[width = \columnwidth]{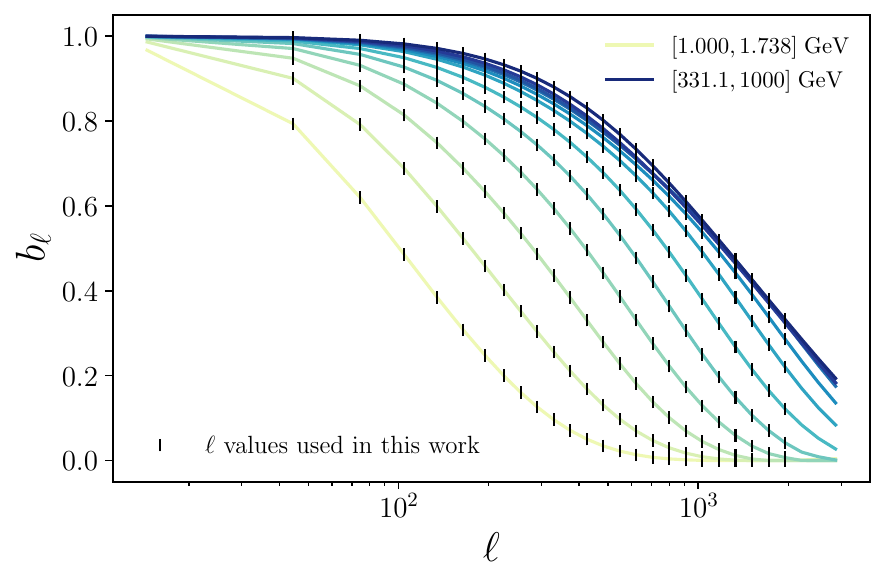}
        \caption{The point-spread function (PSF) in harmonic space $b_{\ell}$ for the 12 \Fermi energy bins considered in this work. The vertical markers indicate the mean of the multipole bandpowers used in this work.}
        \label{fig:fermibeam}
    \end{figure}
    
    We impose masks that remove all bright \gammaray sources detected by {\it Fermi}-LAT and recorded in the LAT 12-year Source Catalog (4FGL-DR3)\footnote{\url{https://heasarc.gsfc.nasa.gov/W3Browse/fermi/fermilpsc.html}}, comprising of more than 6000 sources. The masks are different in each energy bin due to the energy-dependence of the {\it Fermi} point-source response function, which is summarised as follows. The reconstruction of the direction of photons detected by {\it Fermi} is imperfect, and the uncertainty depends strongly on energy. This effectively leads to a smearing of the \gammaray maps, which can be quantified by the PSF. In real space, the PSF  corresponds to the probability of measuring an incoming photon direction that differs from its true direction by an angle $\theta$. This leads to a suppression of power in the \gammaray maps on scales smaller than the typical extent of the PSF. In harmonic space, this is quantified by the harmonic transform of the real-space PSF, which we plot in Fig. \ref{fig:fermibeam}. We see that $b_{\ell}$ is consistently 1 for a larger range of $\ell$ values at higher energy bins than lower bins. In particular, there is significant suppression on medium-large $\ell$ scales, with relatively low energy bins (energies 0.52 GeV to 15 GeV) suffering the most from this effect. We include the PSF in our fiducial analysis by multiplying all theoretical power spectra templates by the corresponding PSF function in harmonic space.

    As well as masking all point sources, we mask the region of the sky for which the fiducial galactic diffuse emission template (gll\_iem\_v07) exceeds three times the isotropic template. Repeating our analysis using a factor four instead, we verified that the results presented here are unaffected by this choice. The isotropic template is spatially constant, with a spectrum given by the \textit{Fermi} Isotropic Spectral Template\footnote{\url{https://fermi.gsfc.nasa.gov/ssc/data/access/lat/BackgroundModels.html}}. 
    The galactic template is calibrated using a model of inverse Compton emission and spectral line surveys of HI and CO and infrared tracers of dust column density and is described in more detail in \citet{1602.07246}.
    
    After applying these masks, we refit the amplitudes of the isotropic and galactic templates in each of the 100 energy bins, maximising a Poisson likelihood. The residuals to this fit are corrected for the \textit{Fermi} exposure and then summed into 12 logarithmically spaced energy bins. The edges of these bins are listed in the second column of Table \ref{tab:Fmeasure}. These summed residual maps are the maps used in the remainder of the analysis.

  \subsection{2MPZ and WISC}\label{ssec:data.gals}
    \begin{figure}
        \centering
        \includegraphics[width = \columnwidth]{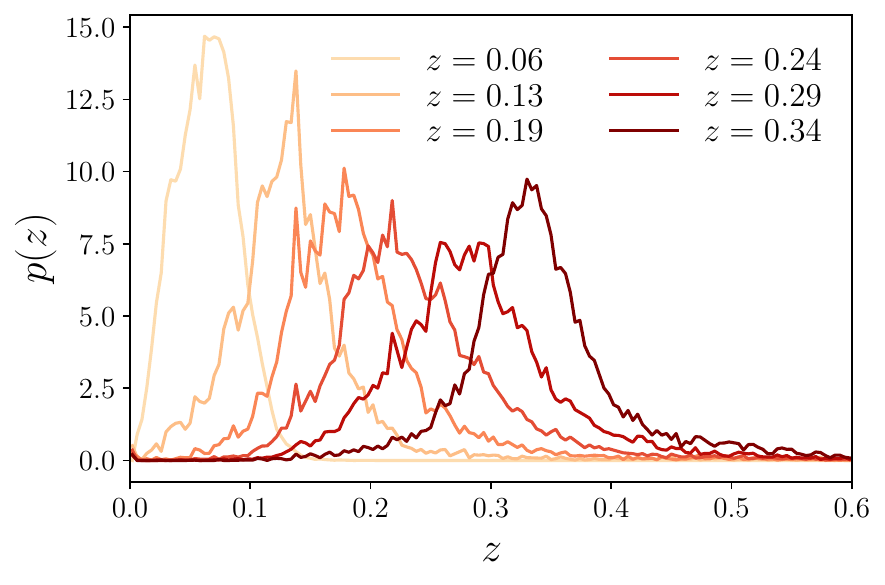}
        \caption{The redshift distributions of the six redshift bins considered in this work. The mean redshift of each respective bin is tabulated with the plot.}
        \label{fig:galkernel}
    \end{figure}
    We make use of photometric galaxy samples from the 2MASS photometric redshift survey (2MPZ \citep{1311.5246}) and the WISE-SuperCOSMOS photometric survey (\wisc \citep{1607.01182}). 2MPZ combines optical and infrared photometry from the Two-Metre All Sky Survey (2MASS \cite{2006AJ....131.1163S,astro-ph/0004318}), the SuperCOSMOS sky survey \citep{astro-ph/0108290,1607.01189}, and the Wide-field Infrared Survey Explorer (WISE \cite{abs/1008.0031}), to obtain high-accuracy photometric redshifts ($\sigma_z\sim0.015$) for over 940{,}000 sources at redshifts $z\lesssim0.15$. \wisc, in turn, sacrifices the 2MASS photometric data (and the corresponding photometric redshift accuracy -- $\sigma_z\sim0.033$) for deeper redshift coverage, comprising about 20 million galaxies at redshifts $z\lesssim0.4$.

    When analysing the 2MPZ and \wisc data, we follow closely the treatment of \cite{1805.11525,1909.09102}. We divide the sample into 6 redshift bins, with 2MPZ comprising the lowest bin, and \wisc being split into 5 bins with equal photometric redshift width ($\Delta z_{\rm ph}=0.05$) in the range $0.1\leq z_{\rm ph}\leq 0.35$. These are the same redshift bins used in \cite{1805.11525,1909.09102}, and further details about the resulting galaxy samples are provided in those references. Unlike in \cite{1805.11525,1909.09102}, we characterised the redshift distribution of each of the six redshift bins employed using the direct calibration (DIR) method of \cite{0801.3822}, using a cross-matched spectroscopic sample including sources from SDSS DR14 \cite{1707.09322}, 2dFGRS \cite{astro-ph/0306581}, WiggleZ \cite{1910.08284}, GAMA \cite{2203.08539}, SHELS \cite{1405.7704}, VIPERS \cite{1708.00026}, and AGES \cite{1110.4371}. Details of the method can be found in \cite{2210.13434}. Galaxy weights were found through a search of the 20 nearest neighbors to each spectroscopic source in the multi-dimensional space of observed magnitudes (8 dimensions for 2MPZ, 4 dimensions for \wisc). The resulting redshift distributions are shown in Fig. \ref{fig:galkernel}.

    Galaxy overdensity maps were constructed from the number counts of galaxies in pixels, and corrected for contamination from extinction and stars as described in \cite{1909.09102}. The residual contamination is limited to large scales ($\ell\lesssim10$), which we remove from our analysis (see Section \ref{ssec:meth.cls}). The associated sky mask was constructed as described in \cite{1909.09102}.
    
\section{Methods}\label{sec:meth}
  \subsection{Power spectrum estimation}\label{ssec:meth.cls}
    To compute all power spectra we use the MASTER algorithm \cite{astro-ph/0105302} implemented in {\tt Namaster}\footnote{\url{https://github.com/LSSTDESC/NaMaster}} \citep{1809.09603}. The estimator can be summarised as follows (refer to \citep{1809.09603} for further details). Observations of a given field are usually on an incomplete sky. We can relate the true map of the field $u(\unitn)$, to the observed map $\tilde{u}(\unitn)$ via $\tilde{u}(\unitn) = g_{u}(\unitn)u(\unitn)$, where, in the simplest case, $g_{u}(\unitn)$ is a binary mask such that $g_{u}(\unitn) = 1$ if a pixel at $\unitn$ has been observed and $g_{u}(\unitn) = 0$ otherwise. In general, $g_u$ may be understood as a local weight, which can be tuned to optimise the precision of the estimated power spectra. To compute the $C_{\ell}$ in terms of the observed maps, we use the ``pseudo-$C_{\ell}$" estimator:
    \begin{equation}\label{eq: meth.pseudocl}
        \tilde{C}^{u v}_{\ell} \equiv \frac{1}{2\ell +1} \sum_{m = -\ell}^{\ell} \tilde{u}_{\ell m} \tilde{v}^{*}_{\ell m},
    \end{equation}
    where $ \tilde{u}_{\ell m}$ and $\tilde{v}^{*}_{\ell m}$ are the harmonic coefficients of the masked maps $\tilde{u}(\unitn)$ and $\tilde{v}(\unitn)$ respectively. Multiplying a true map by a fixed sky mask results in statistical coupling between the different harmonic coefficients, which leads to coupling between different power spectrum multipoles, thus making the pseudo-$C_{\ell}$ estimator biased with respect to the true $C_{\ell}$:
    \begin{equation}
        \langle \tilde{C}^{u v}_{\ell} \rangle  = \sum_{\ell '}M^{g_{u} g_{v} }_{\ell \ell '}C^{u v}_{\ell},
    \end{equation}
    where $\langle \cdots\rangle$ is an ensemble average. $M^{g_{u} g_{v} }_{\ell \ell '}$ is the ``mode-coupling matrix" (MCM), which can be computed entirely in terms of the pseudo-$C_\ell$ of the masks. In principle, we would invert the MCM to obtain an unbiased estimator of the true $C_\ell$. In general, the MCM is not guaranteed to be invertible but can be approximately inverted by binning the pseudo-$C_\ell$ into bandpowers. The procedure is summarised as follows:
    \begin{enumerate}
        \item We bin the pseudo-$C_\ell$ to obtain the mode-coupled bandpowers $\tilde{C}_{q}^{u v} = \sum_{\ell \in q}B^{\ell}_{q}\tilde{C}_{\ell}^{u v}$ and the binned MCM $\mathcal{M}^{g_{u} g_{v}}_{q q'}= \sum_{\ell \in q}\sum_{\ell ' \in q'}B^{\ell}_{q}M^{g_{u} g_{v} }_{\ell \ell '}$, where $B_q$ is a binning operator for the $q$-th bandpower.
        \item We invert the binned MCM to obtain the decoupled, bandpowers $\hat{C}^{u v}_{q} = \sum_{q ' }(\mathcal{M}^{g_{u} g_{v}})^{-1}_{q q'}\tilde{C}^{u v}_{q'}$ 
    \end{enumerate}

    In principle, in order to relate our estimated power spectra to their theoretical prediction, we would need to apply the same binning operation to the latter. This can be achieved by convolving it with the so-called ``bandpower window functions'', which encode the effects of mode-coupling, binning, and MCM-inversion:
    \begin{equation}
      {\cal F}^{g_ug_v}_{q\ell}=\sum_{q'\ell'}({\cal M}^{g_ug_v})^{-1}_{qq'} B^{\ell'}_{q'} M^{g_ug_v}_{\ell'\ell}.
    \end{equation}
    In practice, since the power spectra under study are noisy and rather flat, we can simply approximate 
    \begin{equation}
      C^{uv}_q\equiv \sum_\ell {\cal F}^{g_ug_v}_{q\ell}C^{uv}_\ell\simeq C^{uv}_{\ell_q},
    \end{equation}
    where $\ell_q$ is the central multipole in band $q  $. We use a mixed binning scheme, consisting of linear bin widths of $\Delta\ell = 30$ between $0 \le \ell \le 240$
    and then logarithmic bins until $\ell_{\rm max} = 3071$ with $ \Delta {\rm log}_{10}(\ell) = 0.055$.
    
    To avoid any potential systematics in the galaxy overdensity maps on large angular scales (e.g. extinction or star contamination), we remove the first bandpower ($\ell<30$). To avoid numerical inaccuracies in the spherical harmonic transforms (see Appendix A in \cite{1306.0005}) we limit the smallest scale to $\ell=2N_{\rm side}=2048$. This leaves a total of 24 bandpowers for each cross-correlation. The total data vector, containing $6\times 12$ cross-correlations, thus has 1728 elements.  

    To estimate the statistical uncertainties of these power spectra, we made use of the analytical approach outlined in \cite{1906.11765}. The method assumes that all fields involved are Gaussian-distributed, and accurately accounts for the impact of mode-coupling caused by the presence of sky masks. The method requires an estimate of the power spectra of all fields involved. For this, we use the pseudo-$C_\ell$ estimate corrected by the sky fraction of the masks involved:
    \begin{equation}
      C^{ab,{\rm Cov}}_\ell\simeq\frac{\tilde{C}^{ab}_\ell}{\langle w_aw_b\rangle}.
    \end{equation}
    Here $C^{ab,{\rm Cov}}_\ell$ is the power spectrum between fields $a$ and $b$ used to estimate the covariance matrix, $\tilde{C}_\ell^{ab}$ is the corresponding pseudo-$C_\ell$, $w_a$ is the mask of field $a$, and $\langle\cdots\rangle$ denotes averaging over all pixels in the sky. \cite{1906.11765} found that using this recipe was able to accurately account for the impact of mode-coupling in the covariance matrix. To test the validity of the analytical covariance, we compared it with the statistical uncertainties found via jackknife resampling in some of the cross-correlations, obtaining a reasonable agreement between both approaches.

  \subsection{Galaxy bias modelling}\label{ssec:meth.bias}
    \begin{table}
      \centering
      \caption{Main properties of the 6 galaxy samples used in the analysis: mean redshift (second column), linear bias (third column), and HOD parameters (fourth and fifth columns).}
      \begin{tabular}[t]{ccccc}
      \toprule
      Bin & $\langle z\rangle$ & $b_g$ & $\log_{10}(M_{\rm min}/M_\odot)$ & $\log_{10}(M_1/M_\odot)$ \\  [0.5ex]
      \hline
      \addlinespace[1pt]
      1 & 0.06 & 1.18 & 12.1 & 13.3 \\
      2 & 0.13 & 1.10 & 11.7 & 13.0 \\
      3 & 0.19 & 1.15 & 11.7 & 12.9 \\
      4 & 0.24 & 1.19 & 11.5 & 12.5 \\
      5 & 0.29 & 1.20 & 11.5 & 12.6 \\
      6 & 0.34 & 1.46 & 12.1 & 12.9 \\
      \hline
      \end{tabular}\label{tab:zbins}
    \end{table}
    Our fiducial analysis assumes a linear galaxy bias relation. We fix the linear bias parameter of each redshift bin to the values measured by \cite{1805.11525,1909.09102}, corrected for the fiducial cosmology used here. Given the large uncertainties of the cross-correlations analysed in this work, propagating the small statistical error in the linear bias measurement, obtained from the measurements of the galaxy auto-correlation, has a negligible effect on our final constraints.
    
    For the same reasons, we expect the simplifying assumption of a linear bias relation to be sufficiently accurate for the analysis presented here. To test this, we repeat our analysis assuming a more sophisticated model of the galaxy-halo connection, in the form of a Halo Occupation Distribution (HOD) parametrisation. In particular, we use the HOD model of \cite{astro-ph/0408564}, as implemented in \cite{1909.09102}, for the analysis of the same galaxy samples used here. In this model, the number of galaxies in haloes of a given mass is determined in terms of two free parameters:
    \begin{itemize}
      \item $M_{\rm min}$: the halo mass for which the mean number of central galaxies is 0.5.
      \item $M_1$: the typical mass of haloes hosting one satellite galaxy.
    \end{itemize}
    All other parameters of the model (described in detail in Appendix \ref{app:halo.hod}) were fixed to the values used in \cite{1909.09102}. As we will show, the simpler linear bias parametrisation recovers results that are compatible with those found using this HOD model given the measurement uncertainties. The values of the linear galaxy bias, as well as the best-fit values of $(M_{\rm min},M_1)$, found in \cite{1909.09102} for each redshift bin, are shown in Table \ref{tab:zbins}.

  \subsection{Likelihood analysis}\label{ssec:meth.like}
    The models described in Section \ref{sec:th} depend on a set of primary parameters $\vec{\theta}$. These are either the dark matter parameters $F_n$, describing the function $F(\epsilon)$ (see Sections \ref{sssec:th.gamma_DM.imaps} and \ref{sssec:th.gamma_DM.F}), or the astrophysical parameters $\{\emiss{\epsilon},C_{1h}^{g,E}\}$ (see Section \ref{ssec:th.gamma_astro}). We constrain these parameters from our data vector ${\bf d}$, consisting of all $6\times12$ galaxy-$\gamma$-ray cross-correlations. To do so, we assume that ${\bf d}$ follows a Gaussian likelihood
    \begin{equation}\label{eq:like}
      \chi^2\equiv-2\log p({\bf d}|\vec{\theta})=({\bf d}-{\bf t}(\vec{\theta}))^T\,{\sf C}^{-1}\,({\bf d}-{\bf t}(\vec{\theta}))+K,
    \end{equation}
    where ${\bf t}$ is the theory prediction for our data vector, ${\sf C}$ is the covariance matrix, and $K$ is a normalisation constant.

    When the theory prediction is a linear function of the model parameters, which is the case for both $F_n$ and $\{\emiss{\epsilon},C_{1h}^{g,i}\}$, and assuming wide, uniform priors (as we will do here), the posterior distribution is Gaussian in those parameters by construction, and their mean and covariance can be calculated analytically. Specifically, consider the case:
    \begin{equation}
      {\bf t}={\sf T}\,\vec{\theta},
    \end{equation}
    where ${\sf T}$ is an model-independent $N_d\times N_p$ matrix (with $N_d$ and $N_p$ equal to the number of data points and free parameters, respectively). The posterior mean of $\vec{\theta}$ (which coincides with its maximum a-posteriori -- MAP), and its covariance are simply given by:
    \begin{equation}\label{eq:linreg}
      \vec{\theta}_{\rm MAP}=\left({\sf T}^T{\sf C}^{-1}{\sf T}\right)^{-1}{\sf T}^T{\sf C}^{-1}{\bf d},\hspace{12pt}
      {\rm Cov}(\vec{\theta})=\left({\sf T}^T{\sf C}^{-1}{\sf T}\right)^{-1}.
    \end{equation}

    When constraining $F(\epsilon)$, $\vec{\theta}\equiv\{F_n\}$, and the elements of ${\sf T}$ are (see Eqs. \ref{eq:F_parametrisation} and \ref{eq:cl_parametrisation})
    \begin{equation}
      T^{(g,i,\ell)}_n\equiv C_\ell^{g,i,n}.
    \end{equation}
    When constraining the astrophysical parameters, we can write the parameter vector as $\theta_{\alpha,g,i}$, with $\alpha\in\{1,2\}$, and
    \begin{equation}
      \theta_{1,g,i}\equiv\emiss{E_i(1+z_g)},\hspace{12pt}
      \theta_{2,g,i}\equiv C_{\rm 1h}^{g,i}.
    \end{equation}
    The elements of ${\sf T}$, in turn, are (see Eqs. \ref{eq:model_astro} and \ref{eq:Q_astro}):
    \begin{equation}
      T^{(g,i,\ell)}_{(1,g',j)}\equiv \delta_{ij}\delta_{gg'}Q_\ell^g,
      \hspace{12pt}T^{(g,i,\ell)}_{(2,g',j)}\equiv\delta_{ij}\delta_{gg'}.
    \end{equation}

    After having measured these linear model parameters from a given set of power spectra, we make use of these measurements to constrain other secondary parameters (e.g. DM annihilation cross section or decay rate, or the functional dependence of the $\gamma$-ray emissivity on energy and redshift). In those cases, we still use a Gaussian likelihood, as in Eq. \ref{eq:like}, using the measured linear parameters as data (i.e. with ${\bf d}$ and ${\sf C}$ given by $\vec{\theta}_{\rm MAP}$ and ${\rm Cov}_\theta$ in Eq. \ref{eq:linreg} for the linear parameters). This is mathematically consistent, since the linear dependence of the theory on these primary parameters ensures that they follow a Gaussian distribution. Since these secondary parameters are in general not linearly related to the primary ones, in this case we make use of Markov-Chain Monte-Carlo (MCMC) techniques in order to obtain parameter constraints. For this, we use the {\tt emcee} package\footnote{\url{https://github.com/dfm/emcee}.} \cite{2013PASP..125..306F}.

    All theoretical predictions were computed with the Core Cosmology Library (CCL \cite{1812.05995}), making use of the CAMB Boltzmann solver \cite{astro-ph/9911177} to compute the linear matter power spectrum. The non-linear power spectrum, where necessary, was computed using the {\tt HALOFit} implementation of \cite{1208.2701}. We fixed all cosmological parameters to the best-fit values found by {\sl Planck} \cite{1807.06209}. In halo model calculations, we made use of the halo mass function parametrisation of \cite{0803.2706}, the halo bias model of \cite{1001.3162}, and the concentration-mass relation of \cite{0804.2486}. We use a spherical overdensity halo mass definition with overdensity parameter $\Delta=200$ with respect to the critical density (see Eq. \ref{eq:th.haloM}). We verified that, changing the mass function and concentration parametrisations (to that of \cite{1502.07357} and \cite{1112.5479}, respectively) leads to only small changes in our results, at the level of 10-20\%, in line with the usual accuracy of most halo model ingredients \cite{1104.0949}.

\section{Results}\label{sec:res}
  \subsection{Power spectrum measurements}\label{ssec:res.cls}
    \begin{figure}
        \centering
        \includegraphics[width = \columnwidth]{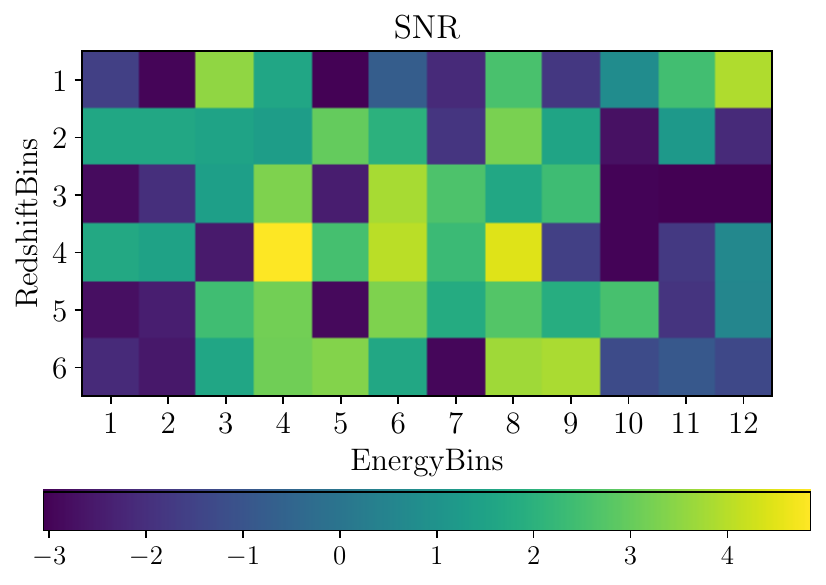}
        \caption{The signal-to-noise ratio ($\widetilde{\rm SNR}$) estimates given a null model of each respective 6 $\times$ 12  galaxy-clustering and \gammaray cross-correlations. The negative values arise from the sign function in Eq. \ref{eq:snr}}
        \label{fig:snr}
    \end{figure}
    \begin{figure*}
        \centering
        \includegraphics[width = 0.9\textwidth]{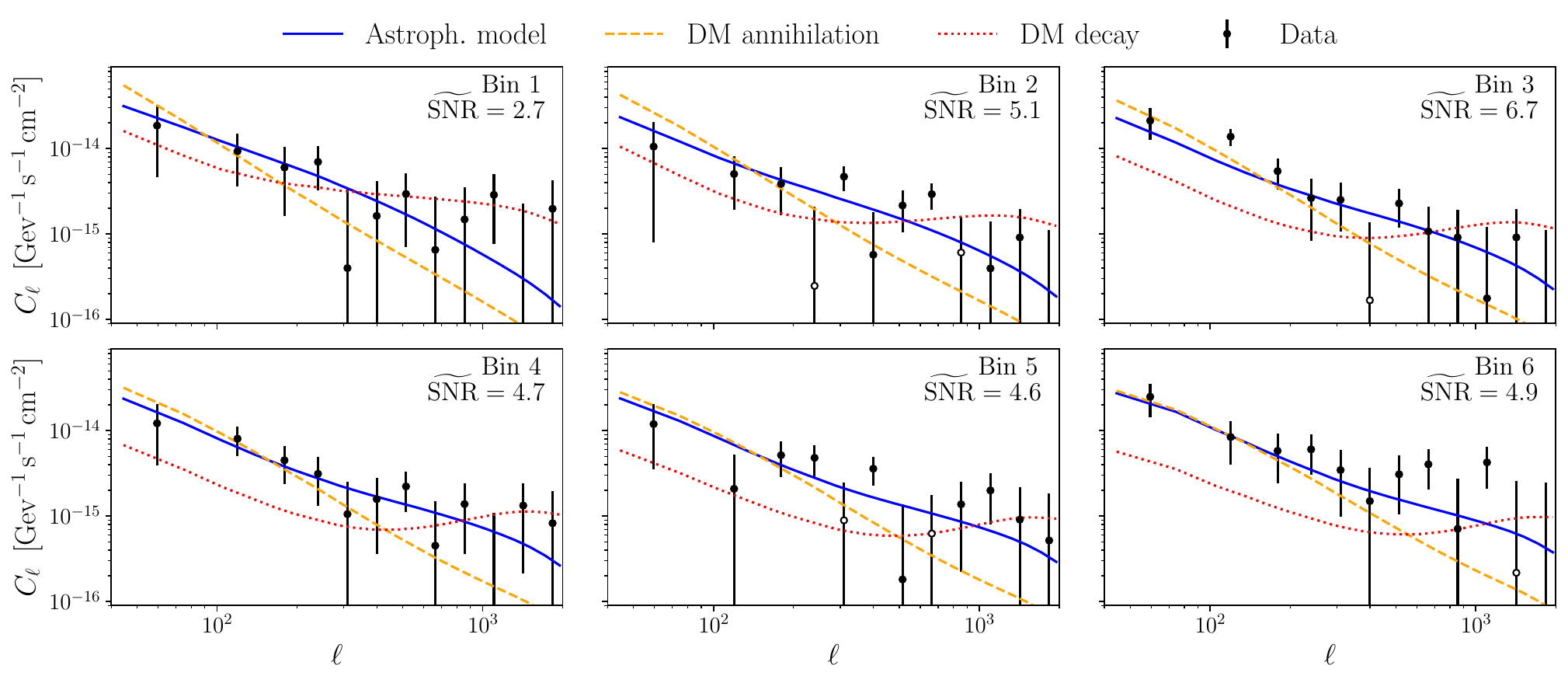}
        \caption{Cross-correlations between \Fermi and the 6 2MPZ and \wisc galaxy samples. Each panel shows the power spectrum coadded over \gammaray energies assuming a power-law spectrum with spectral index $\alpha=-2.3$ and inverse-variance weighting. The measured power spectra are shown as black circles with error bars. Empty circles show the absolute value of a given measurement when negative. The solid blue lines show the best-fit predictions for the astrophysical model presented in Section \ref{ssec:res.astro}, while the dashed orange, and dotted red lines show the best predictions for DM annihilation and decay, respectively, described in Section \ref{sssec:res.dm.F} (obtained from the model-independent reconstruction of $F(\epsilon)$, and hence independent of the specific decay/annihilation channel and WIMP mass).}
        \label{fig:cls}
    \end{figure*}

    We compute the angular power spectrum for the galaxy-clustering maps in 6 redshift bins with the \gammaray maps in 12 energy bins, resulting in 72 cross-correlations in total. We shall refer to an individual galaxy-\gammaray cross-correlation with the notation: {\it redshift bin $\times$ energy bin}.

    As a preliminary exploration of the data vector, we compute a rough estimate of the signal-to-noise ratio (SNR) in each power spectrum as
    \begin{equation}\label{eq:snr}
      \widetilde{\rm SNR}\equiv{\rm sign}(\chi^{2}_0 - N_d)\sqrt{|\chi^{2}_0 - N_d|},
    \end{equation}
    where $N_d=24$ is the number of data points in any given power spectrum, and $\chi_0^2$ is the $\chi^2$ statistic (see Eq. \ref{eq:like}) for a null model (${\bf t}=0$). Since we are not fitting the data to a particular model at this stage (we do not consider any noise models), we shall take Eq. \ref{eq:snr} as a crude estimate of the detection significance for our cross-correlations. The rationale behind Eq. \ref{eq:snr} is that the expectation value of the $\chi^2$ for purely noise-dominated Gaussian noise is $N_d$, and thus the equation is a measure of the departure with respect to this expectation. Note that we will use a more principled definition for SNR when adopting a given model in the rest of the paper.

    In Fig. \ref{fig:snr}, we present a visualisation of $\widetilde{\rm SNR}$ for the $6\times12$ power spectra analysed in this work. From this qualitative estimate, we observe that the signal measurement is dominated by the intermediate energy bins (3 GeV $-$120 GeV), and that the 2-3 highest- and lowest-energy bins are largely noise-dominated. This is as expected, as \Fermi is most sensitive at intermediate energies \cite{0902.1089}. The highest $\widetilde{\rm SNR}$ is achieved in the $4 \times 4$ power spectrum, with a tentative $\sim 5\sigma$ detection of the signal.

    Since each of the 72 power spectra explored carries only a relatively small part to the total SNR of the data, determining whether a signal is being consistently detected is not straightforward by analysing any individual spectrum. As we will see in Section \ref{ssec:res.astro}, the data supports a model in which the $\gamma$-ray emissivity scales with rest-frame energy approximately as $\epsilon^\alpha$, with $\alpha\sim-2.3$. We can use this to produce measurements of the cross-correlation between galaxy redshift bin $g$ coadded over all 12 energy bins as follows:
    \begin{enumerate}
      \item We multiply the cross-correlation with the $i$-th energy bin by the inverse of the energy scaling, $(\varepsilon_{i,g}/\varepsilon_*)^{-\alpha}$, where $\varepsilon_{i,g}\equiv E_i(1+z_g)$ is a rough estimate of the mean rest-frame energy in each bin, with $E_i$ the mean energy of the $i$-th bin, and $z_g$ the mean redshift of the $g$-th galaxy sample. We use a pivot energy $\epsilon_*\equiv20\,{\rm GeV}$, and the best-fit value of $\alpha = -2.3$ found in Section \ref{ssec:res.astro}.
      \item We correct this cross-correlation by the beam transfer function for that energy bin (see Fig. \ref{fig:fermibeam}).
      \item The result of the two previous steps is a set of 12 power spectra that, assuming perfect correlation across energies, should roughly correspond to the same quantity: the cross-correlation between galaxies and the $\gamma$-ray emissivity at $\epsilon=\epsilon_*$. The coadded $C_\ell$ is then computed as the inverse-variance-weighted mean of these 12 rescaled spectra.
    \end{enumerate}
    Note that we only do this here for visualisation purposes. All steps of our analysis in the next sections are based on the raw set of $6\times12$ power spectra.
    
    Fig. \ref{fig:cls} shows the six coadded cross-correlations, as well as the approximate $\widetilde{\rm SNR}$ of each bin. We see that we obtain consistently positive cross-correlations, detected between $2.7\sigma$ and $6.7\sigma$. The figure also shows the best-fit theoretical prediction for the astrophysical model described in Section \ref{ssec:th.gamma_astro} (solid blue line), and for the DM decay and annihilation models of Section \ref{ssec:th.gamma_DM} (dotted red and dashed orange lines). These results are presented in more detail in Sections \ref{ssec:res.dm} and \ref{ssec:res.astro}. Although all models are able to provide a reasonably good fit to the data, these preliminary results already show that the DM models are not flexible enough to fully reproduce the redshift evolution of the signal. We will quantify this more accurately in Sections \ref{sssec:res.dm.F} and \ref{ssec:res.astro}.
        \begin{figure*}
            \centering
            \includegraphics[width = 0.8\textwidth]{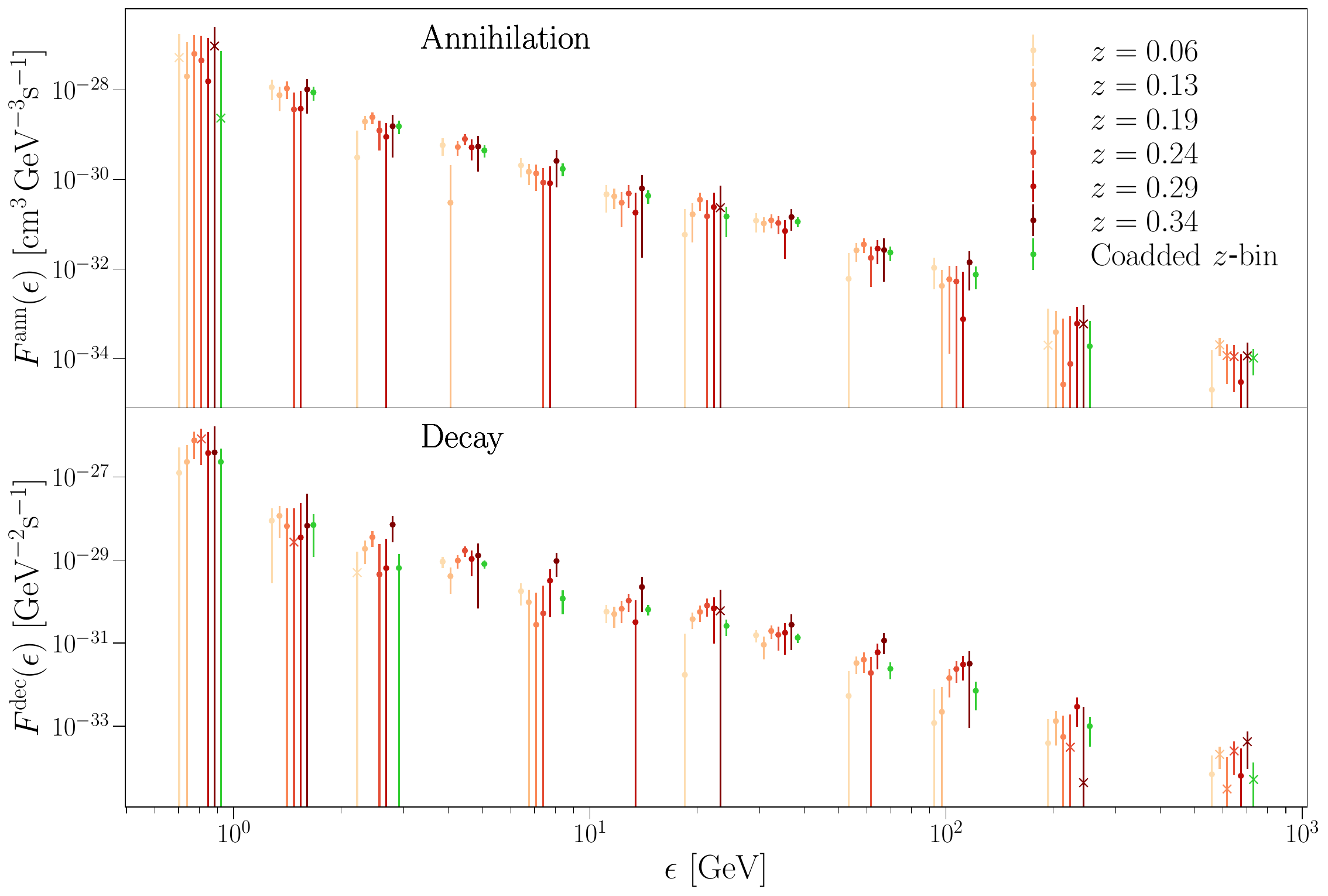}
            \caption{The tomographic and coadded measurements of $F_n$ for annihilation (top panel) and decay (bottom panel). The horizontal shift in data points is for visualisation purposes. The points marked by a cross are the negative values of $F_n$, and the points marked by a dot are the positive values.}\label{fig:Feps}
        \end{figure*}

    Using the best-fit theoretical predictions for the three models explored in the next sections, we can produce more accurate estimates of the detection significance of the signal, given by
    \begin{equation}
      {\rm SNR}_M\equiv\sqrt{\chi^2_0-\chi^2_M-N_\theta+1}
    \end{equation}
    where, as before, $\chi^2_0$ is the $\chi^2$ with respect to a null model, $\chi^2_M$ is the $\chi^2$ with respect to the best-fit prediction within model $M$, and $N_\theta$ is the number of free parameters of the model. Using this definition, together with the best-fit models presented in the next sections (DM decay and annihilation from the model-independent reconstruction of $F(\epsilon)$, and astrophysical sources), we obtain:
    \begin{align}
      {\rm SNR}_{\rm decay}= 8.2,\hspace{12pt}
      {\rm SNR}_{\rm ann.}=8.6,\hspace{12pt}
      {\rm SNR}_{\rm astro}= 9.7.
    \end{align}
    We thus find that the cross-correlation between the diffuse $\gamma$-ray emission as measured by Fermi, and the 2MPZ $+$ \wisc galaxy samples is detected at the $\sim8$-10$\sigma$ level.

  \subsection{Dark matter constraints}\label{ssec:res.dm}
    \begin{table*}
        \centering %
        \caption{Coadded measurements of $F_n$ for annihilation and decay, for each of the 12 \gammaray energy bins weighted by the energy spectrum with an index of $\alpha = -2.3$, together with their 68\% uncertainties. The annihilation constraints are presented for the 4 different models of substructure described at the end of Section \ref{sssec:th.gamma_DM.imaps} and in Appendix \ref{app:halo.ann}.} 
        \begin{tabular}[t]{cccccc}
        \toprule 
        $E\,\,[{\rm GeV}]$ & $F^{\rm ann}_{\rm G12} [{\rm cm}^{3}{\rm GeV}^{-3}s^{-1}]$ &$F^{\rm ann}_{\rm M16} [{\rm cm}^{3}{\rm GeV}^{-3}s^{-1}]$ & $F^{\rm ann}_{\rm SC14} [{\rm cm}^{3}{\rm GeV}^{-3}s^{-1}]$  & $F^{\rm dec} [{\rm GeV}^{-2}s^{-1}]$  \\  [0.5ex]
         \hline
         \addlinespace[1pt]
         $0.71$ & $(-6.23 \pm 37.7) \times 10^{-29}$ & $(-2.34 \pm 71.1) \times 10^{-29}$ & $(-2.43 \pm 76.2) \times 10^{-29}$ &  $(2.32 \pm 2.54) \times 10^-27$ \\
         $1.29$ & $(3.82 \pm 1.47) \times 10^{-29}$ & $(8.79 \pm 3.14) \times 10^{-29}$ & $(9.42 \pm 3.38) \times 10^{-29}$ & $(7.02 \pm 5.82) \times 10^-29$ \\
         $2.25$&  $(5.62 \pm 2.2) \times 10^{-30}$ & $(1.54 \pm 0.496) \times 10^{-29}$ & $(1.66 \pm 0.535) \times 10^{-29}$ & $(6.47 \pm 7.53) \times 10^-30$ \\
         $3.91$ &$(2.51 \pm 0.557) \times 10^{-30}$ & $(4.45 \pm 1.31) \times 10^{-30}$  & $(4.68 \pm 1.42) \times 10^{-30}$  &  $(8.1 \pm 1.8) \times 10^-30$ \\
         $6.52$ & $(6.56 \pm 2.15) \times 10^{-31}$ & $(1.72 \pm 0.543) \times 10^{-30}$ & $(1.86 \pm 0.588) \times 10^{-30}$ &  $(1.19 \pm 0.686) \times 10^-30$ \\
         $11.20$ &  $(2.12 \pm 0.574) \times 10^{-31}$ & $(4.32 \pm 1.48) \times 10^{-31}$ & $(4.57 \pm 1.6) \times 10^{-31}$ & $(6.42 \pm 1.84) \times 10^-31$ \\
         $18.90$ &  $(7.25 \pm 3.49) \times 10^{-32}$ & $(1.5 \pm 0.988) \times 10^{-31}$ & $(1.59 \pm 1.08) \times 10^{-31}$ &  $(2.61 \pm 1.13) \times 10^-31$ \\
         $29.65$ &  $(4.93 \pm 1.07) \times 10^{-32}$ & $(1.14 \pm 0.279) \times 10^{-31}$ & $(1.21 \pm 0.303) \times 10^{-31}$ &  $(1.36 \pm 0.352) \times 10^-31$ \\
         $53.59$ & $(8.18 \pm 3.22) \times 10^{-33}$ & $(2.34 \pm 0.849) \times 10^{-32}$ & $(2.52 \pm 0.921) \times 10^{-32}$ & $(2.42 \pm 1.06) \times 10^-32$ \\
         $94.28$ &  $(3.28 \pm 1.44) \times 10^{-33}$ & $(7.52 \pm 3.94) \times 10^{-33}$ & $(8.02 \pm 4.28) \times 10^{-33}$ &  $(7.16 \pm 4.71) \times 10^-33$ \\
         $187.64$ &  $(2.15 \pm 2.07) \times 10^{-34}$  & $(1.89 \pm 5.14) \times 10^{-34}$ & $(1.79 \pm 5.55) \times 10^{-34}$ & $(1.01 \pm 0.685) \times 10^-33$ \\
         $534.99$  & $(-2.69 \pm 2.43) \times 10^{-35}$ &$(-1.03 \pm 0.603) \times 10^{-34}$ & $(-1.13 \pm 0.652) \times 10^{-34}$ & $(-5.23 \pm 8.04) \times 10^-35$ \\
        \hline
        \end{tabular}
        \label{tab:Fmeasure}
    \end{table*}

    \subsubsection{Measuring $F$}\label{sssec:res.dm.F}        
      We produce measurements of the step-wise amplitudes $F_n$ characterising the energy dependence of the model-independent quantity $F(\epsilon)$ (see Section \ref{sssec:th.gamma_DM.F}) from our power spectrum measurements, using the analytical solution for linear parameters described in Section \ref{ssec:meth.like}. We do so for each redshift bin individually (combining all cross-correlations with the 12 energy bins), and for the full data vector containing all $6\times12$ cross-correlations.

      The results are shown in Fig. \ref{fig:Feps} for annihilation (top panel) and decay (bottom panel). We can see that, within each redshift bin, as well as for the coadded measurements of $F_n$, the function $F(\epsilon)$ seems to be well described by a power-law behaviour. Focusing on the energy bins with highest detection significance (e.g. $\epsilon\simeq10\,\,{\rm GeV}$), we further observe a coherent evolution of the signal with redshift, with the amplitude of $F$ growing with $z$. Since, by construction, $F(\epsilon)$ depends only on particle-physics quantities, and should therefore not be redshift-dependent, this already provides qualitative evidence that the observed signal is not consistent with a purely DM origin.

      \begin{table}
        \caption{Constraints on the free parameters of the power-law model for $F(\epsilon)$ (see Eq. \ref{eq:plawF}) for DM annihilation and decay, in each of the 6 redshift bins (first 6 rows), and for the coadded measurements (last row).}
        \centering
        \begin{tabular}[t]{ccccc}
        \toprule 
        $\langle z \rangle$ & $F^{\rm ann}_0\times10^{30}$ & $F^{\rm dec}_0\times10^{30}$ & $-\alpha^{\rm ann}$ & $-\alpha^{\rm dec}$ \\[0.5ex]
        & $[{\rm cm}^{3}{\rm GeV}^{-3}s^{-1}]$ & $[{\rm GeV}^{-2}s^{-1}]$ & & \\
         \hline
         \addlinespace[1pt]
         0.06 & $0.149 \pm 0.047$ & $0.137 \pm 0.038$ & $2.32 \pm 0.19$ & $2.24 \pm 0.16$ \\
         0.13 & $0.116 \pm 0.031$ & $0.138 \pm 0.036$ & $2.37 \pm 0.16$ & $2.28 \pm 0.15$ \\
         0.19 & $0.163 \pm 0.032$ & $0.237 \pm 0.046$ & $2.36 \pm 0.13$ & $2.22 \pm 0.12$ \\
         0.24 & $0.144 \pm 0.032$ & $0.245 \pm 0.055$ & $2.37 \pm 0.13$ & $2.27 \pm 0.12$ \\
         0.29 & $0.126 \pm 0.033$& $0.302 \pm 0.070$ & $2.20 \pm 0.16$ & $2.06 \pm 0.12$ \\
         0.34 & $0.153 \pm 0.039$ & $0.430 \pm 0.105$ & $2.41 \pm 0.16$ & $2.35 \pm 0.15$ \\
         Coadd & $0.154 \pm 0.020$ & $0.174 \pm 0.023$ & $2.31 \pm 0.08$ & $2.19 \pm 0.08$ \\
        \hline
        \end{tabular}
        \label{tab:plaw}
      \end{table}
    
      The coadded measurements of $F_n$ are listed in Table \ref{tab:Fmeasure} for DM decay and annihilation, including all 4 models of substructure explored. Since the model used to parametrise the cross-correlations is fully described by the $F_n$ (having fixed the cosmological and galaxy bias parameters), these measurements compress all the particle physics information contained in our 1728-element data vector into only 12 numbers. Furthermore, since these measurements have been obtained without assuming any specific decay/annihilation channels, they can be used directly to place constraints on arbitrary DM particle physics models. It is worth noting that the uncertainties on the measured $F_n$ in different energy bins are not entirely uncorrelated. The full covariance matrix is made publicly available together with the measurements of $F$.

      We find that our measurements of $F$ are robust to the choice of a simple linear bias parametrisation used here, with results changing by less than 10\% and 40\% of the statistical uncertainties for annihilation and decay, respectively, when using the HOD model described in Section \ref{ssec:meth.bias}.

      As an example, we fit the measured $F_n$ in each redshift bin to a simple power-law model of the form
      \begin{equation}\label{eq:plawF}
        F(\epsilon)=F_0\left(\frac{\epsilon}{\epsilon_0}\right)^\alpha.
      \end{equation}
      We choose a pivot frequency $\epsilon_0=20\,\,{\rm GeV}$, and fit for the amplitude $F_0$ and spectral index $\alpha$ as free parameters. We imposed flat priors on both parameters: $\alpha \in [-3 ,0]$ for both decay and annihilation, and $F_{0}^{\rm dec} \in [0.001, 3] \times 10^{-29}\,\,{\rm GeV^{-2}s^{-1}}$ and $F_{0}^{\rm ann} \in [0.01, 3] \times 10^{-29}\,\,{\rm {\rm cm}^3GeV^{-3}{\rm s}^{-1}}$. We sample the resulting posterior distribution using {\tt emcee}.

      We tabulate the constraints on $F_0$ and $\alpha$ for each redshift bin, and for the coadded measurements of $F_n$ in Table \ref{tab:plaw}. For the coadded measurements, we obtain the following spectral indices: $\alpha^{\rm ann} = -2.31 \pm 0.08$  and $\alpha^{\rm dec} = -2.19 \pm 0.08$ for annihilation and decay respectively. \cite{1812.02079} modelled the unresolved \gammaray spectrum detected by {\it Fermi}-LAT as a double power-law with an exponential cut-off, obtaining the following spectral indices: $-2.55 \pm 0.23$ and $-1.86 \pm 0.15$, where the two spectral indices correspond to the case in which the detected spectrum is sourced by a double-population scenario. As noted in \cite{1812.02079}, these spectral indices are compatible with blazar-like sources and mis-aligned active galactic nuclei at energies above and below a few GeV, respectively. Our measurement for the annihilation and decay spectral indices lie in between both of these measurements, with the annihilation spectral index being compatible with the first measurement. Our measurements lie somewhat in between both values. \cite{2205.12916} obtained a spectral index of $-2.75^{+ 0.71}_{-0.46}$, in agreement with our results. \cite{1002.3603} obtained a slightly steeper index of $-2.41 \pm 0.05$ for a single power-law by modelling detected {\it Fermi}-LAT sources and the diffuse Galactic \gammaray emission for energy ranges of 0.2 to 100 GeV (probing most of the energy range considered in this work). Our annihilation spectral index is compatible with \cite{1002.3603}. Our results are in excellent agreement with those of \cite{1709.01940} (see e.g. their Table 5), who determined the spectral index of diffuse \gammaray emission through cross-correlations of earlier Fermi data with a large set of galaxy samples, including 2MPZ and WI-SC.
        \begin{figure}
            \centering
            \includegraphics[width = \columnwidth]{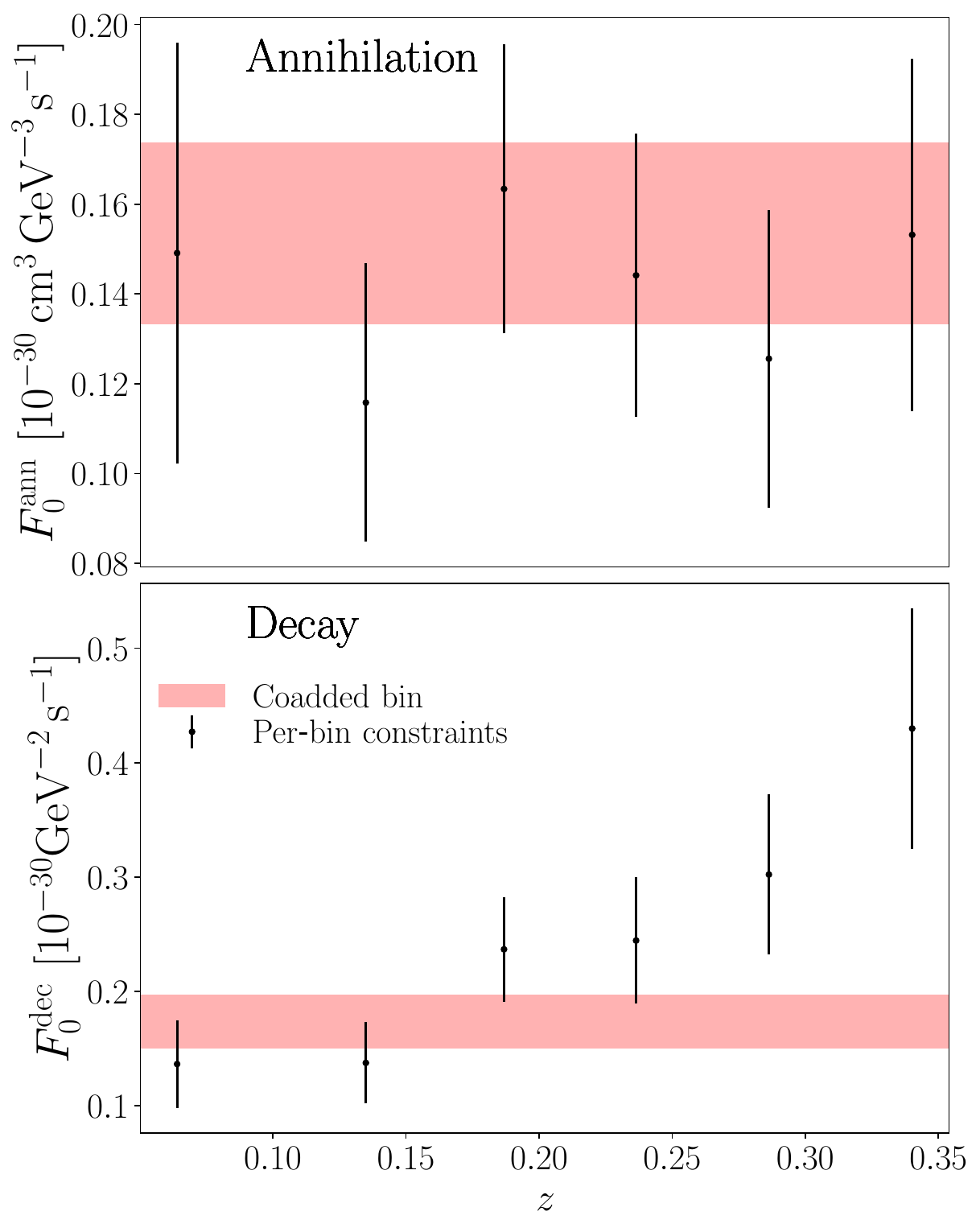}
            \caption{Constraints on the amplitude of the $F(\epsilon)$ function, in the power-law model of Eq. \ref{eq:plawF}, for DM annihilation (top panel) and decay (bottom panel). The points with error bars show constraints in each redshift bin, with the corresponding mean redshift shown in the $x$ axis. The shaded horizontal bands show the constraint obtained from the coadded measurement of $F(\epsilon)$.}
            \label{fig:F0vsz}
        \end{figure}
      
      Our measurements of $F_0$ for decay and annihilation are shown in Fig. \ref{fig:F0vsz} as a function of redshift. As already anticipated in Figs. \ref{fig:cls} and \ref{fig:Feps}, we observe a trend for the amplitude of $F(\epsilon)$ to grow with redshift, especially for DM decay, which should not be the case if the detected signal were caused by DM processes. Quantifying the evidence for this trend is not entirely straightforward a priori, given the non-negligible overlap between the six redshift bins explored here. We will do so in Section \ref{ssec:res.astro}.

      This trend indicates that at least a sizeable fraction of the measured signal is likely caused by astrophysical, baryonic sources, rather than DM processes. Thus, without a reliable prediction for what this fraction is, in what follows we will treat our measurements as contributing to the upper bound on the contribution to the diffuse \gammaray background from DM decay and annihilation. Specifically, when quoting an upper bound on a given DM property $\mu$ (e.g. $\sigv$ of $\Gamma$), we will quote $\bar{\mu}+2\sigma_\mu$, where $\bar{\mu}$ and $\sigma_\mu$ are the mean and standard deviation of the inferred quantity. Note that, since we will fix the WIMP mass when constraining $\Gamma$ and $\sigv$, these parameters are still linearly related to the $F_n$ measurements, and to our original data vector of cross-correlations, and thus their statistical uncertainties are Gaussianly distributed.
      
    \subsubsection{Constraints on DM 
    parameters}\label{sssec:res.dm.pp}
      \begin{figure*}
          \centering
          \includegraphics[width = 0.9\textwidth]{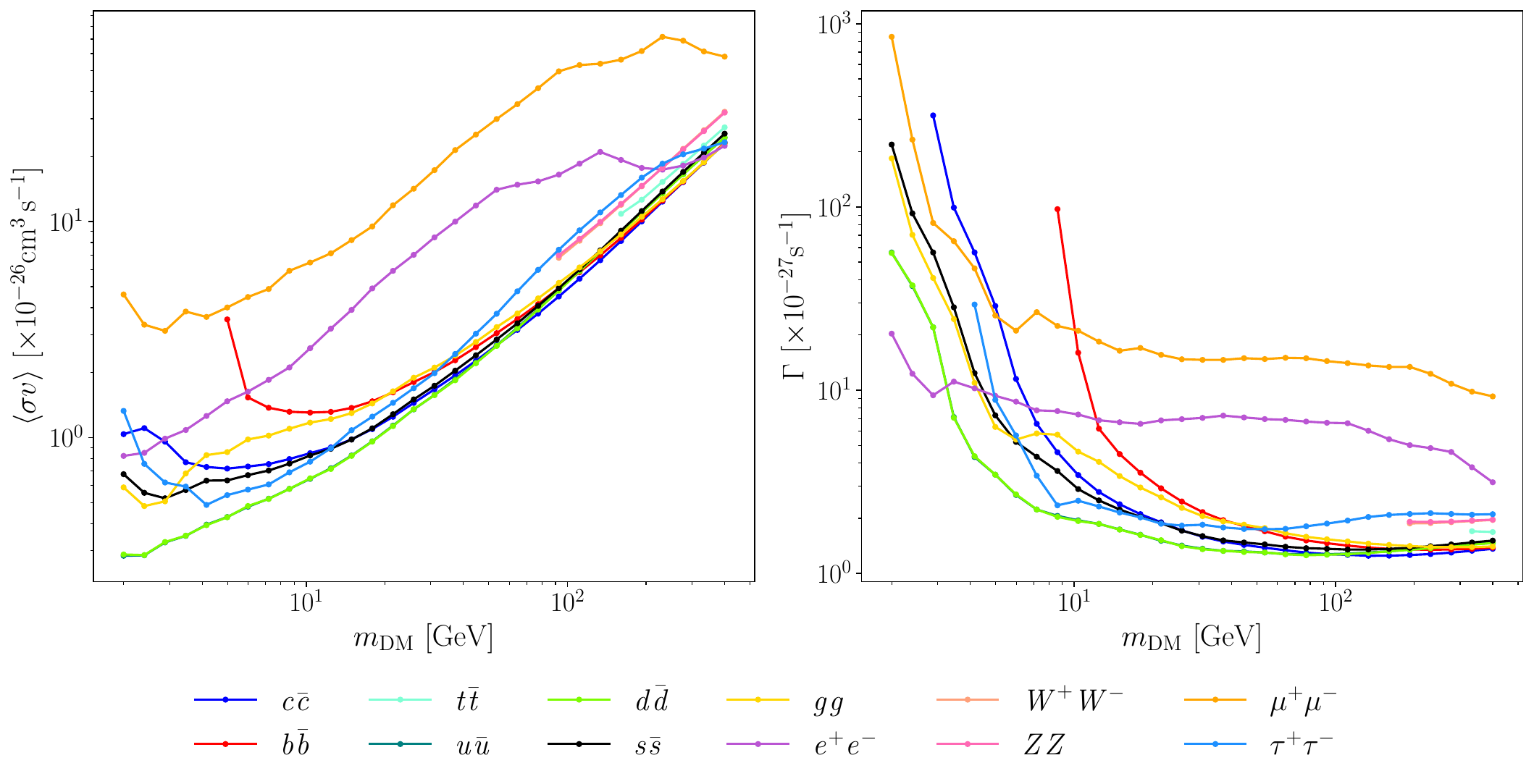}
          \caption{95\% upper bounds on the DM annihilation cross section (left) and decay rate (right) as a function of WIMP mass. The different lines show constraints assuming a single decay/annihilation channel into Standard Model particle-antiparticle pairs (see legend).}
          \label{fig:gammafullplot}
      \end{figure*}
    
    To obtain the constraints on DM annihilation and decay properties, we follow the prescription outlined in Eq. \ref{ssec:meth.like} and employ Eq. \ref{eq:linreg}.
    The data vector ${\bf d}$ is now our measurements $F(\epsilon)$ and we set $\vec{\theta}_{\rm MAP}$ to be the DM quantities we wish to constrain: the velocity-averaged annihilation rate $\sigv$ and the decay rate $\Gamma$ for DM annihilation and decay respectively, as a function of the WIMP mass $m_{\rm DM}$. We set the ${\sf T}$ matrix of the linear regression (Eq. \ref{eq:linreg}) to be (recall Eq. \ref{eq:Fann} and Eq. \ref{eq:Fdec}):
    \begin{equation}
        {\sf T} = 
        \Biggl\{
        \begin{array}{cc}
        \frac{1}{m_{\rm DM}^{2}}\frac{dN}{d\epsilon}, &  {\rm Annihilation}\\
        \frac{1}{m_{\rm DM}}\frac{dN}{d\epsilon}, & {\rm Decay.}
        \end{array} 
    \end{equation}
    To determine ${\sf T}$, we use the {\sl Fermi} Tools and {\tt FermiPy} to obtain the theoretical spectrum $dN/d\epsilon$ for both annihilation and decay. Specifically, we use {\tt DMFitFunction}\footnote{\url{https://fermi.gsfc.nasa.gov/ssc/data/analysis/scitools/source\_models.html}} in {\tt FermiPy} to calculate the spectrum $dN/d\epsilon$ as a function of DM mass, and decay channel. Since the signal follows roughly a power law, with the spectral indices quoted in the previous section, instead of convolving the spectrum with the energy bandpass of each channel, we simply evaluated the function at effective energies weighted by this power-law spectrum:
    \begin{equation}
        \epsilon_{{\rm eval}, i} = \frac{\int^{\epsilon_{i+1}}_{\epsilon_{i}}d\epsilon  \epsilon^{\alpha +1}}{\int^{\epsilon_{i+1}}_{\epsilon_{i}}d\epsilon \epsilon} = \frac{\alpha + 1}{\alpha + 2}\frac{\epsilon^{\alpha + 2}_{i+1}-\epsilon^{\alpha + 2}_{i}}{\epsilon^{\alpha + 1}_{i+1} - \epsilon^{\alpha + 1}_{i}},
    \end{equation}
    using the values of $\alpha$ for decay and annihilation found above. We compute $\sigv$ and $\Gamma$ at fixed DM mass, considering logarithmically spaced values within the range $[2,400]\,\, {\rm GeV}$. 
    
    In Fig. \ref{fig:gammafullplot}, we present the 95$\%$ confidence limit constraints on $\sigv$ and $\Gamma$ for different Standard Model (SM) particle-anti-particle channels.
    The constraints derived here are contingent on the following implicit assumptions/conditions: 
    \begin{enumerate}
        \item The only progenitor of \gammarays are DM particles, i.e we have not taken into account any astrophysical models that could potentially source the \gammarays. 
        \item Annihilation occurs non-relativistically with orbital angular momentum $L = 0$ (so-called $s$-wave annihilation). This allows us to treat the decay spectrum to be equivalent to the annihilation spectrum but with half of the DM mass.
        \item The DM particle decays/annihilates into single SM particle-anti-particle channels with an branching ratio of 1, i.e. we omit the possibility of decaying/annihilating into multiple channels per event. 
        \item We impose kinematic constraints, such that a decay can only occur if $m_{\rm DM} \ge 2m_{\rm SM}$ and similarly $m_{\rm DM} \ge m_{\rm SM}$ for annihilation, where the subscript $_{\rm SM}$ denotes the corresponding SM particle. Hence some particle channels (e.g $t\Bar{t}$) do not probe the full range of DM masses.
    \end{enumerate}
    
    With these assumptions in mind, we find that the constraints derived for both $\Gamma$ and $\sigv$ are competitive with previous studies that have considered cross-correlations between \gammarays and other cosmological tracers: weak-lensing \cite{1404.5503,1607.02187,1611.03554}, galaxy surveys \cite{1503.05922,1506.01030}, as well as constraints derived from local structures \cite{1001.4531,1201.0753,1312.7609,1503.02641,1611.03184}.
    In the following section, we shall only consider the $b\bar{b}$, $\mu^{+}\mu^{-}$ and $\tau^{+}\tau^{-}$ channels for comparative purposes and for sake of clarity. For  $\Gamma$, we find that across all three channels, we obtain constraints that are of the same order of magnitude at $m_{\rm DM}\sim100\,\,{\rm GeV}$ when compared to constraints found by \cite{1503.05922}. For the $b\bar{b}$ channel and at a DM mass of $\sim 10\,\, {\rm GeV}$, we obtain constraints that are less stringent than those of \cite{1611.03554} for their conservative (DM + Astro) and optimisitc (only DM) models. But, as we go towards higher masses, at around $\sim 100\,\,{\rm GeV}$, we obtain constraints that are slightly tighter than both these models. For both the $\mu^{+}\mu^{-}$ and $\tau^{+}\tau^{-}$ channels, our constraints are nearly an order of magnitude tighter for all masses greater than $\sim 10\,\,{\rm GeV}$ than both DM and DM + Astro models. At $\sim 10\,\,{\rm GeV}$, all three channels are roughly at the same order of magnitude as \cite{2205.12916} with the $b\Bar{b}$ constraint derived in this work being weaker by a factor of ~6. The constraints on the $b\bar{b}$ channel are up to two orders of magnitude weaker than those derived from other cross-correlations with galaxy surveys \cite{1506.01030,1503.05922}, with $\mu^{+}\mu^{-}$ and $\tau^{+}\tau^{-}$ channels being rouhgly the same order of magnitude. 
      
      \begin{figure*}
          \centering
          \includegraphics[width = \textwidth]{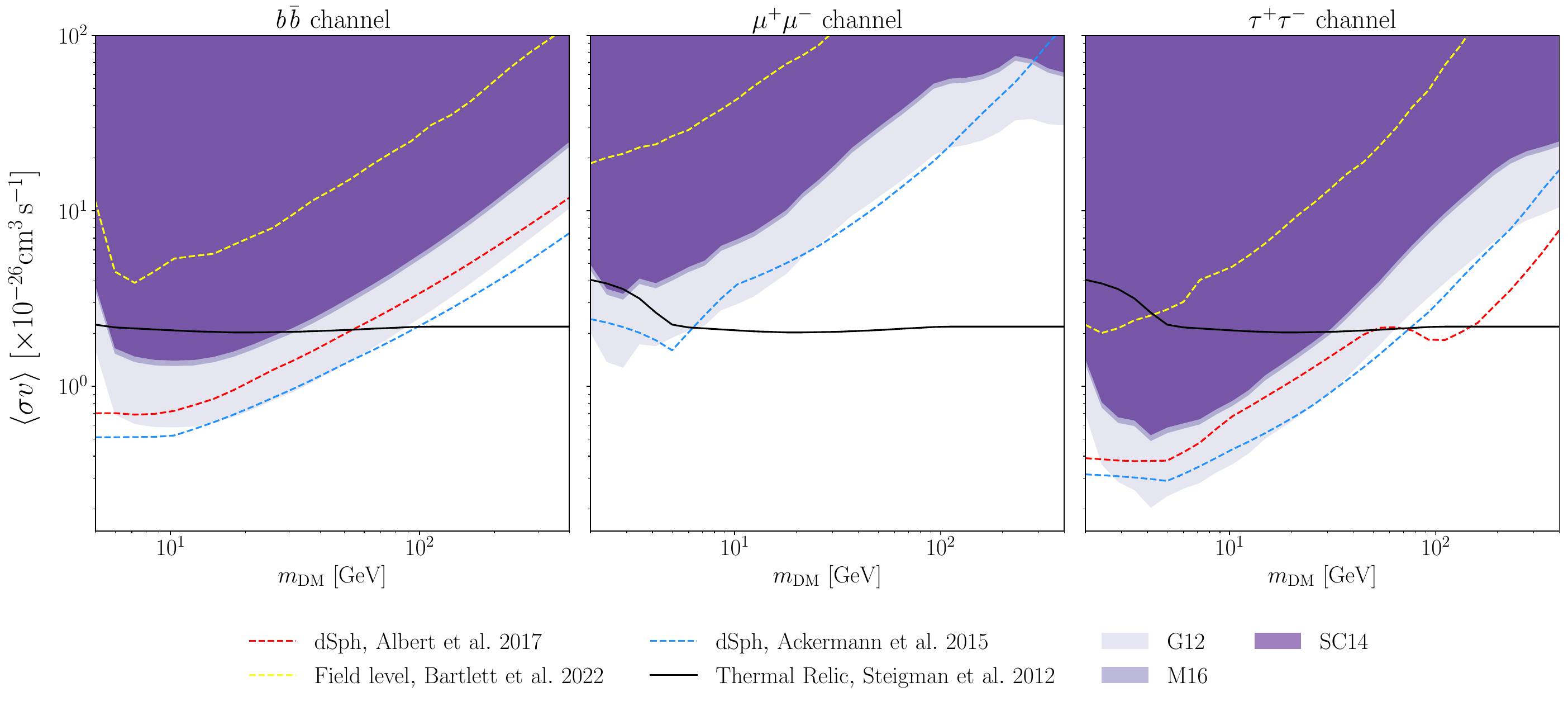}
          \caption{95\% upper bounds on the DM annihilation cross-section for the $b\bar{b}$, $\mu^+\mu^-$, and $\tau^+\tau^-$ channels. The constraints found from our data are shown as shaded band. Constraints found for the four different substructure models described in Appendix \ref{app:halo.ann} are shown as purple bands with different shadings. The yellow dashed line shows the constraints found by \cite{2205.12916} from a field-level analysis of low-redshift structures. The dashed red and blue lines show the constraints obtained from dwarf spheroidals (dSph), as computed by \cite{1611.03184} and \cite{1503.02641}, respectively. The solid black line shows the thermal relic limit \cite{1204.3622}.}\label{fig:chanconstraints}
      \end{figure*}

    Likewise, we find that the constraints on $\sigv$ derived in this work are competitive with those found in the literature. In Fig. \ref{fig:chanconstraints}, we present an exclusion plot for $b\bar{b}$, $\mu^{+}\mu^{-}$ and $\tau^{+}\tau^{-}$ annihilation channels where the shaded areas represent regions that can be ruled out from our 95 $\%$ limit constraints in three substructure regimes: G12, M16, and SC14. The constraints on $\sigv$ across all substructure models have been computed by extrapolating the halo mass down to $10^{-6}\,\,M_{\odot}$ for the sake of concurrence with constraints presented in current literature. We explore the effect of changing this lower bound of the mass in Appendix \ref{app:Mmin}. When comparing equivalent substructure models, we find that the constraints derived in this work are up to roughly two orders of magnitude more stringent than \cite{1611.03554}, obtained from cross-correlation with weak lensing, across all three channels for all three substructure cases. In particular, the G12 constraints obtained are only marginally stronger than both the conservative (DM + Astro) and optimistic (DM only) models for the $b\bar{b}$ channel. In comparison to constraints derived via cross-correlations with galaxy catalogues for equivalent substructure regimes, our constraints are slightly less stringent than \cite{1506.01030} by an order of magnitude at a mass of $\sim 10\,\,{\rm GeV}$.
    This may be because our measurements are limited by the amplitude of the detected astrophysical signal, and hence our upper bound is not determined solely by the statistical error achieved. Although weaker for the $b\Bar{b}$ channel, we find that the constraints derived in this work for both the $\tau^{+}\tau^{-}$ and $\mu^{+}\mu^{-}$ channels for equivalent substructure regimes are up to an order of magnitude more stringent than those found in \cite{1506.01030}. More in detail, our constraints for the $\tau^{+}\tau^{-}$ channel achieves roughly the same constraining power with only slightly tighter constraints in the mass range $10\,\,{\rm GeV} - 100\,\,{\rm GeV}$ for the SC14 substructure regime. 
    For constraints derived through DM-only models \cite{1503.05922}, we find that we can provide roughly the same constraining power for the $b\bar{b}$ channel. We obtain roughly an order of magnitude tighter constraints for $\mu^{+}\mu^{-}$ and $\tau^{+}\tau^{-}$.
    
    It is worth noting, however, that the detailed modelling of the impact of substructure on \gammaray emission from DM annihilation is a large source of theoretical uncertainty in this analysis. In \cite{1503.05922} the difference in the most optimistic case (referred to as LOW in their study, and corresponding to our M16 model) and the most conservative case (HIGH, corresponding to G12) is roughly an order of magnitude. By comparison, we find the difference between these two models to be more modest. Note, however, that, our M16 model \cite{1603.04057} is a refinement of the M16 model used in their work \cite{1312.1729}. The difference between these two cases is a more detailed modelling of halo concentrations. This comparison highlights the sensitivity of the constraints to the substructure modeling parameters.
    The constraints found assuming the potentially optimistic model of G12 for the subhalo boost factor gives rise to a factor of $\sim 3$ difference in the constraints from the SC14 substructure case. The difference in both extreme cases is modest in comparison to \cite{1503.05922}, who report a difference of an order of magnitude. 

    In the case of $b\bar{b}$, we are able to exclude the thermal relic cross section around $m_{\rm DM}\sim10\,{\rm GeV}$, in a range of WIMP masses that depend heavily on the substructure model assumed. In the most optimistic case (G12), for both the $b\bar{b}$ and $\tau^+\tau^-$ channels, the constraints obtained are comparable to those found through the study of dwarf spheroidal galaxies (dSph)  \cite{1503.02641,1611.03184}. Finally, our constraints are more stringent than those found by \cite{2205.12916} from a field-level analysis of low-redshift structures, by a factor that ranges between $\sim2$ and $\sim10$ depending on the substructure model used. 
    \begin{figure*}
        \centering
        \includegraphics[width = 0.7\textwidth]{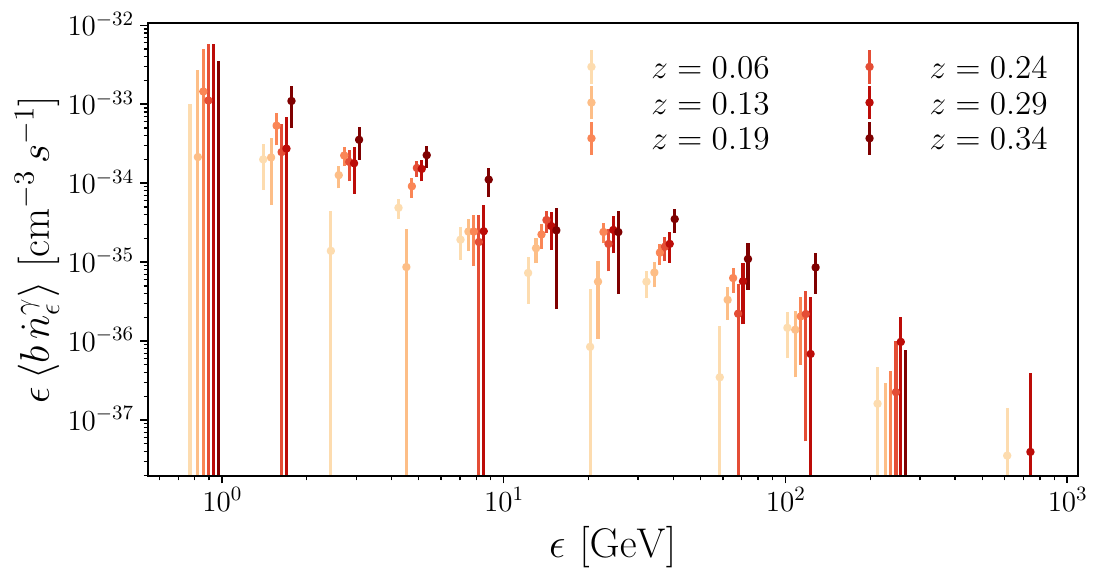}
        \caption{Constraints on the bias-weighted mean \gammaray emissivity as a function of rest-frame energy $\epsilon$ and redshift (see legend). Obtained from the analysis of the galaxy-\Fermi cross-correlations using the model described in Section \ref{ssec:th.gamma_astro}.}
        \label{fig:astro_ez}
    \end{figure*}

  \subsection{Constraints on the diffuse astrophysical \gammaray background}\label{ssec:res.astro}
    In Section \ref{sssec:res.dm.F} we presented evidence that the DM kernel $F(\epsilon)$ evolves with redshift, in a manner that would be incompatible with DM decay or annihilation as the sole origin of the cross-correlation between the diffuse \gammaray background and the positions of galaxies in 2MPZ and WI-SC. This prevents our interpretation of the measured signal in terms of fundamental DM properties, and limits our ability to obtain a tighter upper limit on them.
    
    In this section, we instead take a more agnostic approach, and use our measurements to place constraints on the energy and redshift dependence of the \gammaray emissivity from our data, regardless of its physical origin. To do this, we follow the methodology outlined in Section \ref{ssec:th.gamma_astro}, which allows us to determine the bias-weighted \gammaray emissivity $\emiss{\epsilon}(z)$ from the two-halo regime of the galaxy-\gammaray cross-correlation. Using the linear reconstruction method of Section \ref{ssec:meth.like}, we obtain the measurements shown in Fig. \ref{fig:astro_ez}. It is worth emphasizing that we obtain a single measurement of $\emiss{\epsilon}$ for each individual cross-correlation: the cross-correlation between galaxies at mean redshift $z_g$ and the \gammaray map with mean observed energy $E_i$, provides a measurement of $\emiss{\epsilon}$ at redshift $z_g$ and at the rest-frame energy $\epsilon_{gi}\equiv E_i(1+z_g)$. The position of each measurement along the $x$ axis of the figure corresponds to the associated rest-frame energy (hence the displacement between points at different redshifts).

    \begin{figure}
        \centering
        \includegraphics[width = \columnwidth]{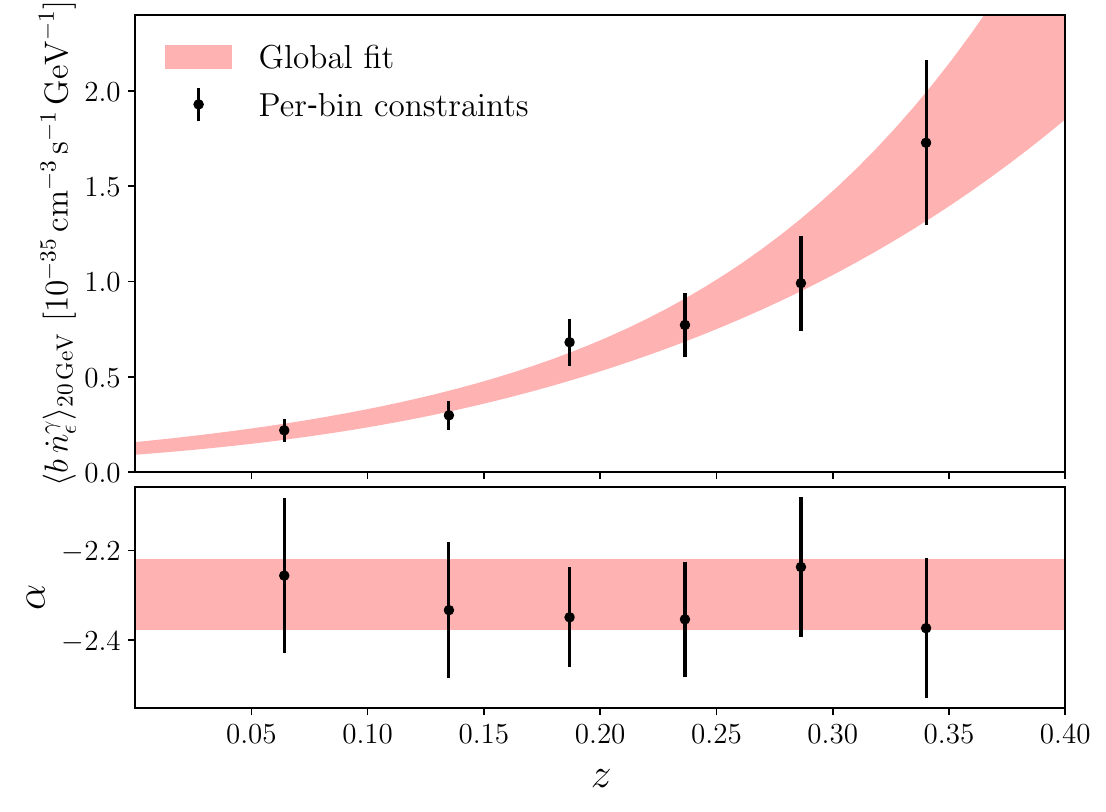}
        \caption{Amplitude (top panel) and spectral index (bottom panel) of the \gammaray emissivity found by fitting the data in Fig. \ref{fig:astro_ez} a power-law model (see Eq. \ref{eq:plaw_e_astro}). Circles with error bars show the constraints from the 6 2MPZ and \wisc redshift bins. The shaded band shows the $1\sigma$ joint constraints found when using the global model of Eq. \ref{eq:plaw_ez_astro}, which additionally accounts for the intrinsic redshift evolution of the emissivity.}
        \label{fig:astro_z_beta}
    \end{figure}
    \begin{table}
      \centering
      \caption{Constraints on the phenomenological parameters of the power-law \gammaray emissivity model (see Eq. \ref{eq:plaw_e_astro}) in each redshift bin.}
      \begin{tabular}[t]{ccccc}
      \toprule
      Bin & $\langle z\rangle$ & $\emiss{\epsilon_0}\,[{\rm cm}^{-3}{\rm s}^{-1}{\rm GeV}^{-1}]$ & $\alpha$ & $\chi^2$ \\  [0.5ex]
      \hline
      \addlinespace[1pt]
      1 & 0.06 & $(2.21\pm0.60)\times10^{-37}$ & $-2.24\pm0.16$ & 10.5 \\
      2 & 0.13 & $(2.94\pm0.77)\times10^{-37}$ & $-2.34\pm0.16$ & 16.9 \\
      3 & 0.19 & $(6.79\pm1.22)\times10^{-37}$ & $-2.35\pm0.11$ & 17.9 \\
      4 & 0.24 & $(7.75\pm1.66)\times10^{-37}$ & $-2.36\pm0.13$ & 13.2 \\
      5 & 0.29 & $(9.88\pm2.49)\times10^{-37}$ & $-2.23\pm0.16$ & 5.2\\
      6 & 0.34 & $(1.73\pm0.43)\times10^{-36}$ & $-2.37\pm0.15$ & 12.1 \\
      \hline
      \end{tabular}\label{tab:astrocon}
    \end{table}
    As in the case of $F(\epsilon)$,  we find that the \gammaray emissivity follows a power-law-like dependence on energy. To quantify this further, we fit the measured values of $\emiss{\epsilon}$ in each redshift bin $g$ to a power-law of the form:
    \begin{equation}\label{eq:plaw_e_astro}
      \emiss{\epsilon}_g=\emiss{\epsilon_0}_g\,\left(\frac{\epsilon}{\epsilon_0}\right)^{\alpha_g},
    \end{equation}
    with $\epsilon_0=20\,{\rm GeV}$ as before. As in Section \ref{sssec:res.dm.F}, since the measured values of $\emiss{E(1+z_g)}$ are linearly related to the cross-power spectra, they also follow a Gaussian likelihood. We assume flat priors for both $\emiss{\epsilon_0}_g$ and $\alpha_g$, and sample the posterior distribution using {\tt emcee}. The results are shown in Table \ref{tab:astrocon}. In all cases, the power-law model is a good fit to the data, with reasonable $\chi^2$ values and associated probabilities. These results are shown as black points with error bars in Fig. \ref{fig:astro_z_beta}. We see that all redshift bins recover compatible values for the spectral index which, as expected, is also in agreement with the spectral index measurement found for $F(\epsilon)$ in Section \ref{sssec:res.dm.F}. Furthermore, there is a clear evolution of the \gammaray emissivity with redshift.

\begin{figure}
        \centering
        \includegraphics[width = \columnwidth]{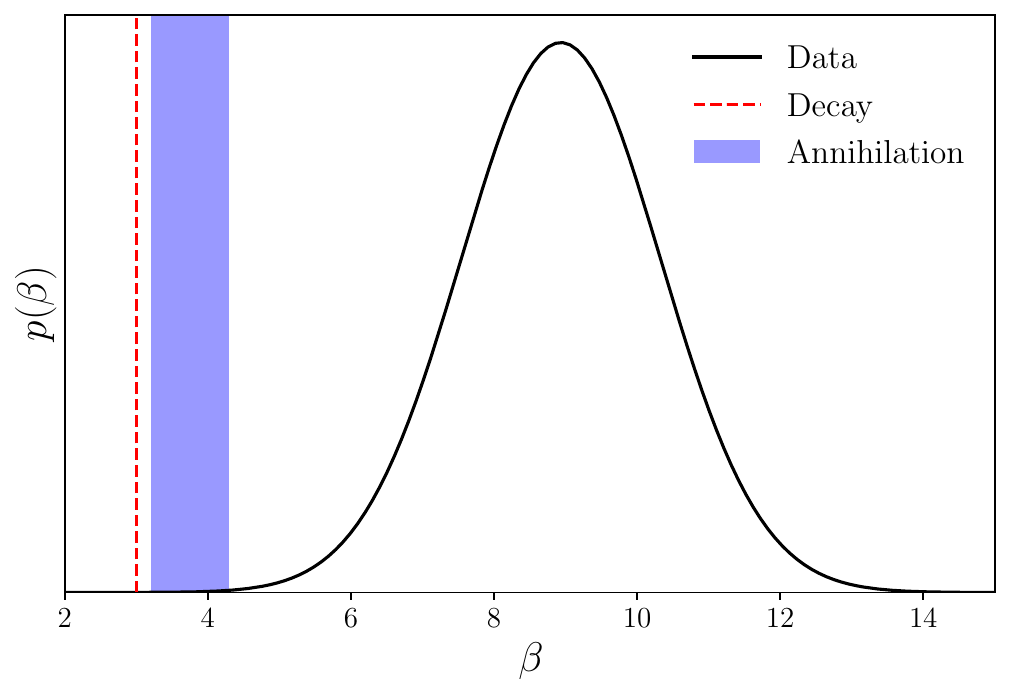}
        \caption{Constraints on the power-law index characterising the redshift evolution of the \gammaray emissivity using the global model in Eq. \ref{eq:plaw_ez_astro}. The black solid line shows the posterior distribution obtained from the set of cross-correlations studied here. The vertical red line and shaded bands show the values of $\beta$ expected for DM decay and annihilation. The measured signal is incompatible with a DM origin at the $\sim3\sigma$ level.}
        \label{fig:astro_eta}
    \end{figure}

    To quantify this redshift evolution, and compare it with the evolution expected for DM decay and annihilation, we fit a global model of the form
    \begin{equation}\label{eq:plaw_ez_astro}
      \emiss{\epsilon}(z)=\emiss{\epsilon_0}_0\left(\frac{\epsilon}{\epsilon_0}\right)^\alpha(1+z)^\beta,
    \end{equation}
    to the values of $\emiss{\epsilon}$ measured from all pairs of redshift and energy bins. The amplitude and spectral index parameters $(\emiss{\epsilon_0}_0 ,\alpha)$ have the same interpretation as before, with the power-law index $\beta$ parametrising the redshift evolution. As before, we use a Gaussian likelihood and assume flat priors for all parameters, obtaining the following constraints:
    \begin{align}\nonumber
      &\emiss{\epsilon_0}_0=(1.22\pm0.35)\times10^{-37}\,\,{\rm cm}^{-3}{\rm s}^{-1}{\rm GeV}^{-1},\\
      &\alpha=-2.30\pm0.08,\hspace{12pt}\beta=8.92\pm1.4.
    \end{align}
    The model has a best-fit $\chi^2$ of 89.5 for 69 degrees of freedom and thus provides an adequate fit to the data. The shaded bands in Fig. \ref{fig:astro_z_beta} show our $1\sigma$ constraints on $\emiss{20\,\,{\rm GeV}}(z)$ for this model, obtained from the MCMC chains. The posterior distribution on the redshift evolution parameter $\beta$ is shown in Fig. \ref{fig:astro_eta}.

    We can use our constraints on $\beta$ to quantify the evidence against a purely DM-related origin for the signal we measured based on its redshift dependence. This is straightforward for DM decay since, in this case, the emissivity is directly proportional to the dark matter overdensity. The value of $\beta$ for DM decay can then be found by simply comparing the radial kernels of both models. The astrophysical kernel is proportional to $(1+z)^{-3}$ (see Eq. \ref{eq:Iint}), whereas the DM kernel is constant (see Eqs. \ref{eq:int_dm_gen} and \ref{eq:dm_dec_C}). Hence, for decay $\beta_{\rm decay}=3$. This value is shown as a vertical red dashed line in Fig. \ref{fig:astro_eta}.
    
    For annihilation, the comparison is less straightforward. The annihilation radial kernel is proportional to $(1+z)^3$ (see Eqs. \ref{eq:int_dm_gen} and \ref{eq:dm_ann_C}). However, annihilation is proportional to $(1+\delta)^2$ instead of $(1+\delta)$, and the power spectrum of both quantities evolves differently with redshift. The effective radial kernel for annihilation is therefore proportional to $(1+z)^{3+\Delta\beta_{\rm ann}}$, and hence $\beta_{\rm ann}\equiv6+\Delta\beta_{\rm ann}$, where
    \begin{equation}\label{eq:delta_beta}
      \Delta\beta_{\rm ann}\equiv\frac{d}{d\log(1+z)}\log\left[\,\frac{P_{\delta,\delta}(k,0)\,P_{\delta,\delta^2}(k,z)}{P_{\delta,\delta}(k,z)\,P_{\delta,\delta^2}(k,0)}\right].
    \end{equation}
    Here $P_{\delta,\delta}(k,z)$ is the power spectrum of the matter overdensity, whereas $P_{\delta,\delta^2}(k,z)$ is the cross-spectrum between $1+\delta$ and $(1+\delta)^2$. The value of $\Delta\beta_{\rm ann}$ depends on scale, as well as on the model used to describe substructure. Over scales relevant for this analysis ($0.01\,{\rm Mpc}^{-1}<k<1\,{\rm Mpc}$), and for the various models of substructure explored here, $\Delta\beta_{\rm ann}$ varies in the range $[-2.8,-1.7]$. The corresponding allowed range of $\beta_{\rm ann}$ is shown as a vertical blue. 
    band in Fig. \ref{fig:astro_eta}.

    Considering the highest value of $\beta$ compatible with DM annihilation, we thus find that our measurements are incompatible with a purely DM-related origin to the diffuse $\gamma$-ray background at the $2.8\sigma$ level (considering the most extreme annihilation case). This justifies our interpretation of the detected signal as an upper bound of the emission from DM processes, rather than an indirect detection of DM decay or annihilation.

\section{Conclusions}\label{sec:conc}
  Observations of the UGRB present a unique probe to improve our understanding of high-energy astrophysics, as well as a window to constrain, or potentially detect, WIMP dark matter. In this work, we have analysed cross-correlations between \gammaray intensity maps, covering the energy range $E\in[0.5\,{\rm GeV},1\,{\rm TeV}]$, with the galaxy overdensity at 6 different redshift ranges covered by the 2MPZ and \wisc surveys.

  We detect a positive cross-correlation between both datasets at the level of $8-10\sigma$, confirming the extragalactic nature of the UGRB. The sensitivity of this measurement, and the availability of redshift data, allows us to reconstruct the dependence of the signal on rest-frame energy and redshift, enabling us to interpret it in the context of both dark matter and astrophysics.

  In the context of WIMP searches, we make use of linear regression methods to reconstruct, from our cross-correlation measurements, a function $F(\epsilon)$, defined in Eqs. \ref{eq:Fann} and \ref{eq:Fdec}, that depends solely on the particle physics parameters governing DM decay and annihilation. We do so in a model-independent way that compresses all the information in our cross-correlations into a set of 12 numbers, $\{F_n\}$, characterising the dependence of this function of rest-frame energy. The reconstructed function has a power-law energy dependence ($F(\epsilon)\propto\epsilon^\alpha$) with a spectral index $\alpha\simeq-2.3$. Repeating this for each redshift bin independently, we find evidence that the amplitude of $F(\epsilon)$ evolves monotonically with redshift. This would not be possible for a signal sourced only by dark matter processes, implying the presence of astrophysical contamination in our measurements. Nevertheless, we may use the measured value of $F(\epsilon)$ to place an upper-bound constraint on the WIMP decay rate and the annihilation cross-section. Although, due to this contamination, the constraints are systematics-limited, we find bounds that are competitive with those obtained by other groups targeting similar large-scale structure cross-correlations, as well as constraints from local structures. As in all other \gammaray WIMP searches, our annihilation constraints are hampered by the theoretical uncertainty on the impact of substructure. Constraints may vary by up to two orders of magnitude between a model with no substructure, and the most extreme substructure model studied here \cite{1312.1729}. In the most optimistic case, we can rule out, at the 95\% confidence level, the thermal relic bound for WIMP masses of a few tens of GeVs assuming complete decay into $b\bar{b}$ quarks or $\tau^+\tau^-$ leptons. Better control over the impact of substructures on the expected annihilation signal must be achieved before such a claim can be undisputed.

  Arguably the main source of uncertainty in our ability to constrain DM physics, is our inability to characterise and clean the contamination from astrophysical sources in the signal, which, as the analysis of Section \ref{ssec:res.astro} shows, must be present. This forces us to interpret the cross-correlation signal, detected at the $\sim8-10\sigma$ level, as contributing to the upper bound on the DM contribution. Together with the uncertainty in the theoretical model from substructures, this constitutes arguably the main impediment in obtaining reliable constraints on DM annihilation. This could be improved in the future by incorporating (and marginalising over) a physics-based model of all potential astrophysical sources. Removing these contaminants at the map level (e.g. by detecting and masking out further extragalactic point sources), would likely lead to larger improvements. Finally, other sectors of the data, including higher-order statistics (e.g. through a field-level analysis of all tracers over the same volume \cite{2205.12916}), as well as including \gammaray auto-correlations and measurements of the isotropic signal, could significantly enhance the constraints presented here. Future \Fermi data (e.g. masking with the new 14-year data source catalog 4FGL-DR4), and its combination with denser and more reliable galaxy catalogs from next-generation surveys, such as DESI \cite{1907.10688}, the Rubin Observatory LSST \cite{1211.0310}, Euclid \cite{1001.0061}, or the Roman Space Telescope \cite{1503.03757}, have the potential to improve the constraints presented here using similar analysis methods, assuming that the modelling challenges outlined above are tackled.

\section*{Acknowledgements}
  We thank Natalia Porqueres for helpful comments on an early version of this work, and Stefano Camera, Nicolao Fornengo, and Tilman Tr\"oster for useful discussions. We also thank the anonymous referee for their comments, which allowed us to improve the quality of this manuscript significantly. AP is partially supported by the National Astronomical Institute of Thailand (NARIT) and St Peter's College, Oxford. DA acknowledges support from the Science and Technology Facilities Council through an Ernest Rutherford Fellowship, grant reference ST/P004474, and from the Beecroft Trust. DJB is supported by the Simons Collaboration on ``Learning the Universe'' and was supported by STFC and Oriel College, Oxford. MB is supported by the Polish National Science Center through grants no. 2020/38/E/ST9/00395, 2018/30/E/ST9/00698, 2018/31/G/ST9/03388 and 2020/39/B/ST9/03494, and by the Polish Ministry of Science and Higher Education through grant DIR/WK/2018/12. We made extensive use of computational resources at the University of Oxford Department of Physics, funded by the John Fell Oxford University Press Research Fund

\bibliography{main}
\appendix
\onecolumngrid
\section{Halo models for the UGRB and galaxies}\label{app:halo}
  \subsection{The mean dark matter profile}\label{app:halo.nfw}
    To describe the mean density of dark matter around haloes, we will use the Navarro-Frenk and White profile (NFW, \citep{astro-ph/9508025}), which takes the form
    \begin{equation}
      \rho_{\rm NFW}(r)=\frac{\rho_0}{x\left(1+x\right)^2}\Theta(r<r_\Delta),\hspace{12pt}x\equiv r/r_s.
    \end{equation}
    Here the comoving spherical overdensity radius $r_\Delta$ is related to the halo mass via
    \begin{equation}\label{eq:th.haloM}
      M=\frac{4\pi}{3}\Delta\,\bar{\rho}_cr_\Delta^3,
    \end{equation}
    where $\bar{\rho}_c$ is the critical density, and $\Delta$ is the so-called spherical-overdensity parameter, which we will set to $\Delta=200$. The characteristic radius $r_s$ is related to $r_\Delta$ through the concentration-mass relation $c(M)$
    \begin{equation}
      r_\Delta=c(M)r_s.
    \end{equation}
    We will use the parametrisation of \cite{0804.2486} to calculate $c(M)$. The normalisation $\rho_0$ can be found by integrating the density profile over volume, and is related to the halo mass via
    \begin{equation}\label{eq:rho0}
      \rho_0=\frac{M}{4\pi\,r_s^3[\log(1+c)-c/(1+c)]}.
    \end{equation}

    The simple form of the NFW profile makes it possible to compute its Fourier transform analytically:
    \begin{align}\nonumber
      \rho_{\rm NFW}(k|M)
      &\equiv4\pi\int_0^\infty dr\,r^2\,\rho_{\rm NFW}(r|M)\,\frac{\sin kr}{kr}\\
      &=\frac{M}{\log(1+c)-c/(1+c)}\,\left[\cos q\left({\rm Ci}((1+c)q)-{\rm Ci}(q)\right)+\sin q\left({\rm Si}((1+c)q)-{\rm Si}(q)\right)-\frac{\sin cq}{1+cq}\right],
    \end{align}
    with $q\equiv kr_\Delta/c$.

  \subsection{The halo occupation distribution}\label{app:halo.hod}
    As an alternative model describing the non-linear relation between the galaxies and matter overdensities, we use the halo occupation distribution framework (HOD, \citep{astro-ph/0109001,1206.6890}). In this case, the number density of galaxies in a halo of mass $M$ is parametrised in terms of the mean number of central and satellite galaxies ($\bar{N}_c(M)$, and $\bar{N}_s(M)$ respectively), as well as the profile describing the distribution of satellites around a given halo $u_s(r|M)$. The mean overdensity of galaxies around a halo of mass $M$ is thus given by
    \begin{equation}
      \langle 1+\delta_g({\bf x}|M)\rangle=\frac{\bar{N}_c(M)}{\bar{n}_g}\left[\delta^D({\bf x})+\bar{N}_s(M)\,u_s(|{\bf x}||M)\right],
    \end{equation}
    where the mean number density of galaxies $\bar{n}_g$ is given by
    \begin{equation}
      \bar{n}_g=\int dM\,n(M)\,\bar{N}_c(M)\left[1+\bar{N}_s(M)\right].
    \end{equation}
    In this work, we will use the same parametrisation employed in \cite{astro-ph/0408564,1909.09102}. The mean number of centrals and satellites is given by:
    \begin{align}
      &\bar{N}_c(M)=\frac{1}{2}\left[1+{\rm erf}\left(\frac{\log(M/M_{\rm min})}{\sigma_{\log M}}\right)\right],\hspace{12pt}
      \bar{N}_s(M)=\Theta(M-M_0)\left(\frac{M-M_0}{M_1}\right)^\alpha.
    \end{align}
    Where $M_{\rm min}$ is the mass at which haloes have on average $0.5$ central galaxies, $M_0$ is the mass at which haloes may start forming satellite galaxies, and $M_1$ is the typical mass for haloes hosting satellites. We further assume that satellites are distributed around haloes following the matter density, and hence
    \begin{equation}
      u_s(r|M)\equiv \rho_{\rm NFW}(r|M)/M.
    \end{equation}

    We will fix the free parameters of the model to the best-fit values for the 2MPZ and \wisc samples found in \cite{1909.09102}. In this case, $M_0=M_{\rm min}$, $\sigma_{\log M}=0$, and $\alpha=1$ in all cases. The remaining free parameters ($M_{\rm min}, M_1$) are fitted in each redshift bin, and take the values found in Table \ref{tab:zbins}.

  \subsection{Halo profile for DM annihilation}\label{app:halo.ann}
    As noted in Section \ref{sssec:th.gamma_DM.imaps}, we consider two limits regarding the impact of substructure on DM annihilation. In the most conservative limit, we study the case with no substructure. In this case, all haloes follow an exact NFW profile, and $\langle \rho_{\rm DM}^2\rangle=\langle\rho_{\rm DM}\rangle^2$. Thus, the halo profile for annihilation without substructure is
    \begin{equation}
      U^{\rm ann,no-sub}_\gamma(r|M)=\frac{\rho_{\rm NFW}^2(r|M)}{\bar{\rho}_{M,0}^2}.
    \end{equation}
    The simplicity of the NFW functional form makes it possible to calculate the Fourier transform of the squared profile analytically:
    \begin{equation}\label{eq:uhost}
      U^{\rm ann,no-sub}_\gamma(k|M)={\cal V}(M)\left[T_1(k|M)+T_2(k|M)+T_3(k|M)+T_4(k|M)\right],
    \end{equation}
    with
    \begin{align}
      &T_1(k|M)\equiv\frac{1}{6q}\left[q(q^2-6)\sin q+3(q^2-2)\cos q\right]\,\left[{\rm Si}((1+c)q)-{\rm Si}(q)\right],\\
      &T_2(k|M)\equiv\frac{1}{6q}\left[q(q^2-6)\cos q-3(q^2-2)\sin q\right]\,\left[{\rm Ci}((1+c)q)-{\rm Ci}(q)\right],\\
      &T_3(k|M)\equiv-\frac{c((q^2-6)c+2q^2-15)+q^2-11}{6q(1+c)^3}\sin(cq),\\
      &T_4(k|M)\equiv\frac{(3c+4)\cos(qc)}{6(1+c)^2}+\frac{{\rm Si}(qc)}{q}-\frac{2}{3},\\
      &{\cal V}(M)\equiv\frac{4\pi r_s^3\rho_0^2}{\bar{\rho}_{M,0}^2},
    \end{align}
    where $\rho_0$ is given in Eq. \ref{eq:rho0}, and $q\equiv kr_\Delta/c$ as before.

    To describe the effects of substructure, we extend this model as outlined in \cite{1301.5901}. In this case, the contribution from subhaloes is characterised by a mass-dependent boost factor $b_{\rm sh}(M)$. The associated halo profile is then given by
    \begin{equation}\label{eq:uannsub}
      U^{\rm ann}_\gamma(k|M)=U^{\rm ann,no-sub}_\gamma(k|M)+b_{\rm sh}(M)\,V_{\rm host}(M)\,u_{\rm sh}(k|M),
    \end{equation}
    where $U^{\rm ann,no-sub}_\gamma(k|M)$, given in Eq. \ref{eq:uhost}, is the smooth contribution from the squared host halo density, and the second term is the contribution from the squared density of subhaloes. This neglects a term of the form ${\rm host}\times{\rm subhalo}$, which is only relevant on scales where the densities of host halo and subhaloes are similar. In Eq. \ref{eq:uannsub}, $V_{\rm host}$ is simply the volume integral of the host profile,
    \begin{equation}
      V_{\rm host}(M)\equiv4\pi \int dr\,r^2\,U^{\rm ann,no-sub}_\gamma(r|M)={\cal V}(M)\left[1-\frac{1}{(1+c)^3}\right],
    \end{equation}
    and $u_{\rm sh}$ is the squared density profile of subhaloes normalised to have a unit volume integral. For this, we use the functional form of \cite{1301.5901} which, in Fourier space, is
    \begin{align}
      u_{\rm sh}(k|M) = A\left[\int_0^1 dx \frac{x^{2}}{(x^{2} + 1/16)^{3/2}}\frac{\sin{\kappa x}}{\kappa x}+ \frac{64}{17^{3/2}}\frac{\kappa \cos{\kappa} + \eta \sin{\kappa}}{\kappa (\kappa^{2} + \eta^{2})}\right],
    \end{align}
    where $A=0.636$ ensures that $u_{\rm sh}(k\rightarrow0|M)\rightarrow1$, and $\kappa\equiv kr_\Delta$.
    \begin{figure}
        \centering
        \includegraphics[width = 0.7\textwidth]{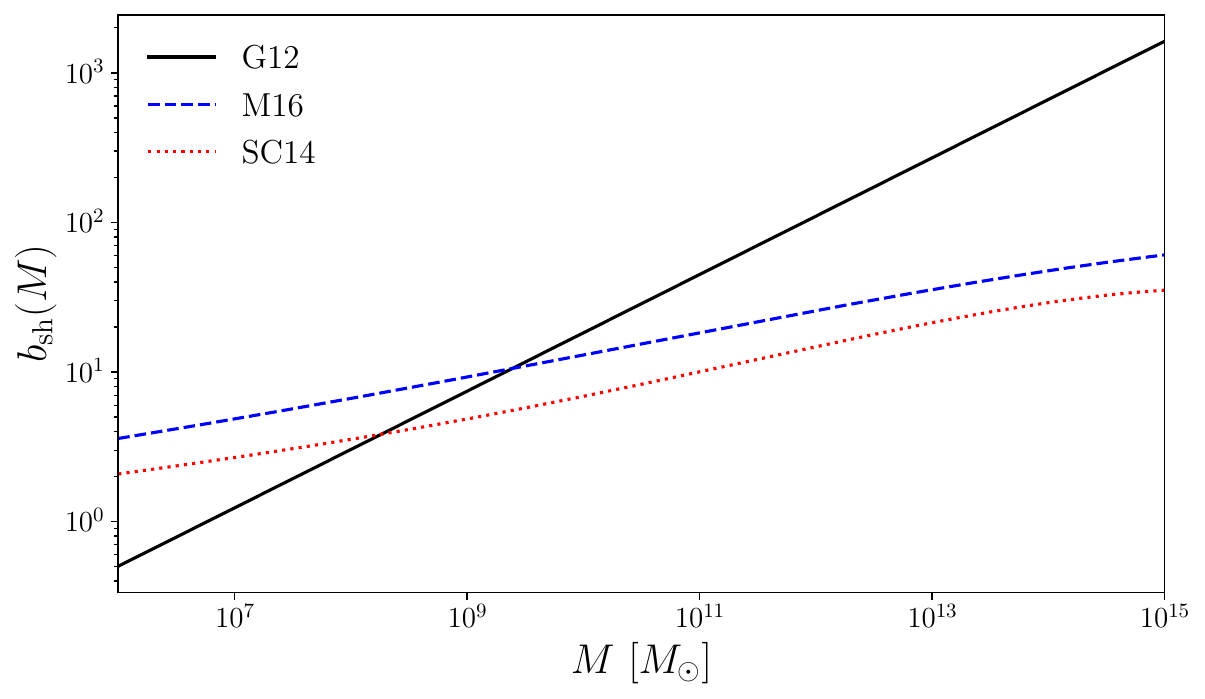}
        \caption{Boost factor as a function of host halo mass for with the three models of substructure used in this analysis, based on \cite{1107.1916} (solid black), \cite{1603.04057} (dashed blue), and \cite{1312.1729} (dotted orange).}
        \label{fig:boost}
    \end{figure}

    The amplitude of the contribution from substructures is controlled by the boost factor $b_{\rm sh}(M)$. We consider three different models of substructure. Ordering them by the amplitude of the effective boost provided to the annihilation signal, these are:
    \begin{itemize}
      \item {\bf SC14}. We use the fitting function of \cite{1312.1729}:
      \begin{equation}
        \log_{10}b_{\rm sh}(M)=\sum_{n=0}^5b_n\log(M)^n,  
      \end{equation}
      with $b_n=\{-0.442,0.0796,-0.0025, 4.77\times10^{-6}, 4.77\times10^{-6},-9.69\times10^{-8}\}$.
      \item {\bf M16}. We use the fitting fucnction of \cite{1603.04057}:
      \begin{equation}
        \log_{10}b_{\rm sh}(M)=\sum_{n=0}^5b_n\log_{10}(M)^n,  
      \end{equation}
      with $b_n=\{-0.186,0.144,-8.8\times10^{-3},1.13\times10^ {-3},-3.7\times10^{-5},-2\times10^{-7}\}$.
      \item {\bf G12}. We use the fitting function of \cite{1107.1916}:
      \begin{equation}
        b_{\rm sh}=110\left(\frac{M}{10^{12}M_\odot}\right)^{0.39}.
      \end{equation}
    \end{itemize}
    The boost factors associated with each of these models are shown in Fig. \ref{fig:boost}. In all cases, the boost factor parametrisations were determined assuming a minimum subhalo mass of $M_{\rm min}=10^{-6}\,M_{\rm min}$.

\section{Minimum halo mass for annihilation}\label{app:Mmin}
  \begin{figure*}
      \centering
      \includegraphics[width=0.49\textwidth]{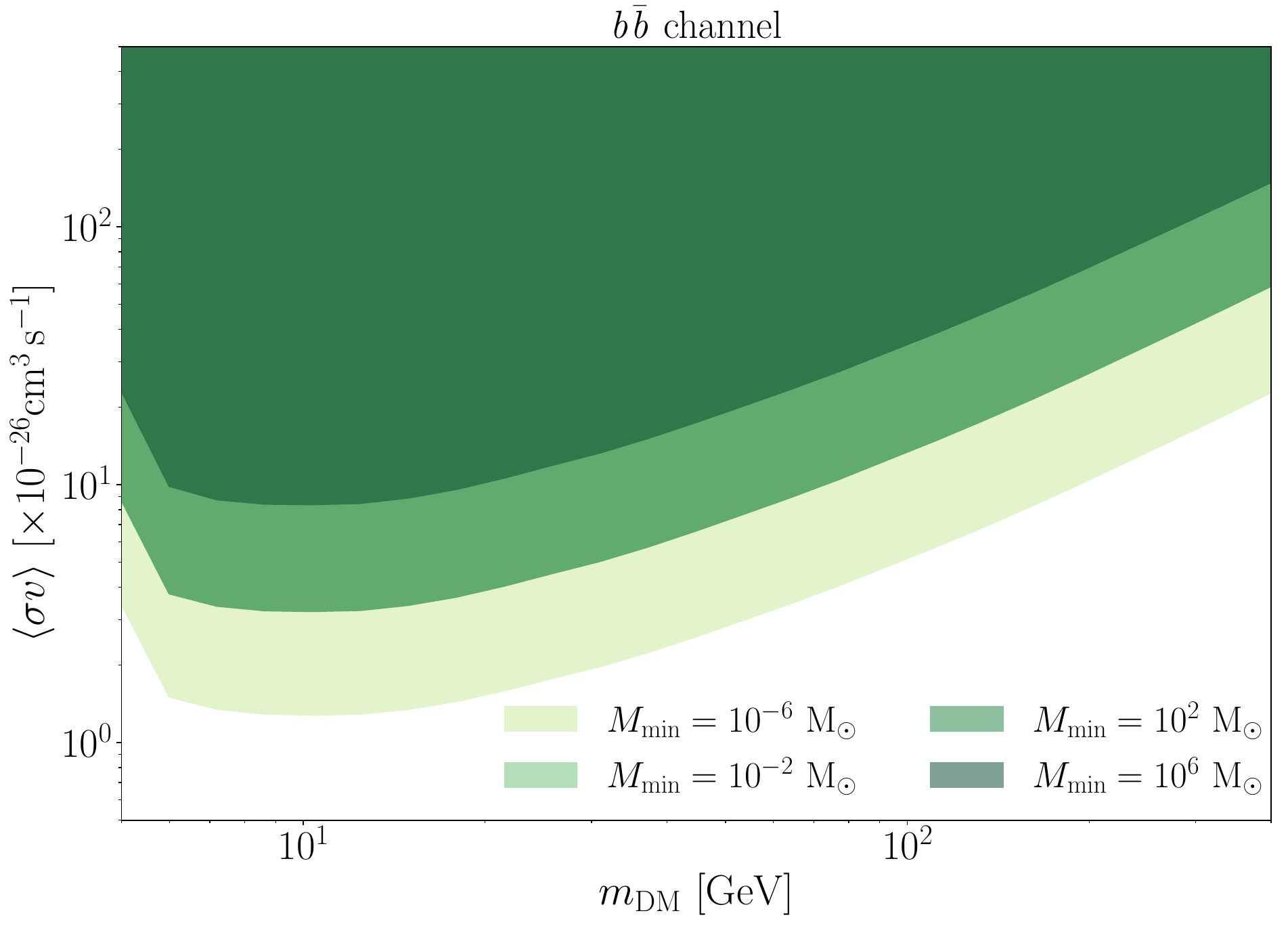}
      \includegraphics[width=0.49\textwidth]{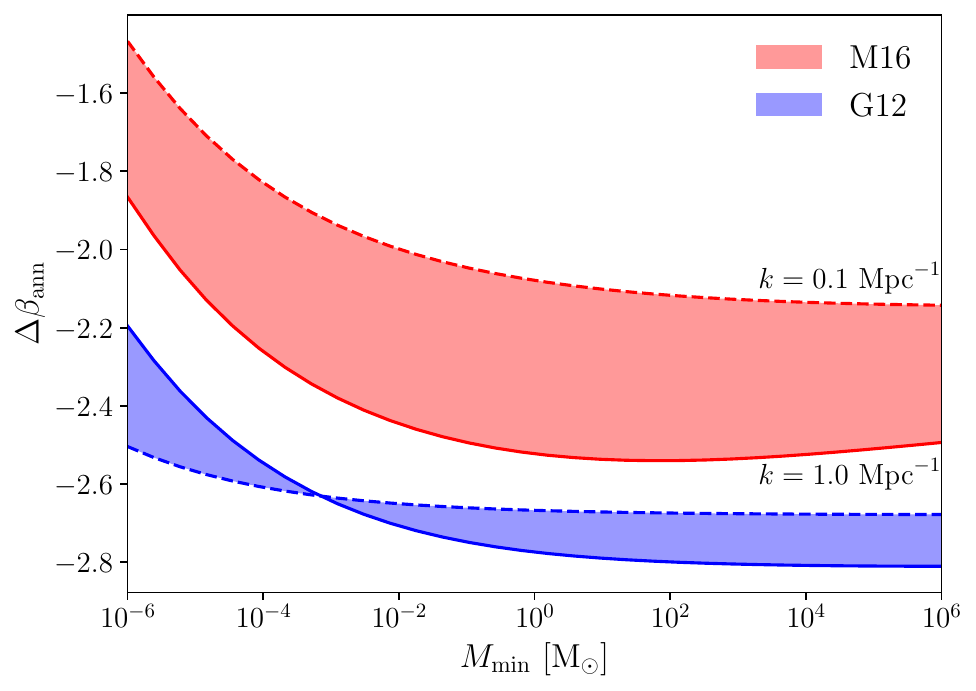}
      \caption{{\sl Left panel:} 95\% upper bounds on the DM annihilation cross section for the M16 substructure case. The increasing darkness of green corresponds to increasing the minimum mass $M_{\rm min}$. The lightest shade corresponds to our fiducial constraints presented in Section \ref{sssec:res.dm.pp}. {\sl Right panel:} the redshift evolution parameter $\Delta\beta_{\rm ann}$ (see Eq. \ref{eq:delta_beta}) as a function of $M_{\rm min}$ for the M16 and G12 substructure models (red and blue bands, respectively).}
      \label{fig:mmin_plot}
  \end{figure*}
  One of the most important sources of theoretical uncertainties when deriving constraints from DM annihilation, is the minimum halo mass contributing to the signal. In this work, we have fixed this to $M_{\rm min}=10^{-6}\,\,M_\odot$, following most other analyses in the literature (e.g. \cite{1506.01030}). Fig. \ref{fig:mmin_plot} shows, however, that the choice of $M_{\rm min}$ can have a significant impact on the $\sigv$ constraints, and on their interpretation.

  The left panel shows the 95\%-level constraints on $\sigv$ obtained assuming different values for $M_{\rm min}$, in the range $[10^{-6}\,M_\odot,10^6\,M_\odot]$. The constraints may vary by up to one order of magnitude depending on this choice. The value of $M_{\rm min}$ also affects the redshift evolution of the signal, and thus can affect the interpretation of tomographic measurements. The right panel shows the redshift evolution parameter $\Delta\beta_{\rm ann}$, defined in Eq. \ref{eq:delta_beta}, as a function of minimum halo mass. Depending on the model of substructure used, $\Delta\beta$ may take values spanning the range $-2.8\lesssim\Delta\beta\lesssim-1.5$. It is worth noting that, although these results do quantify the uncertainty due to the exact minimum halo mass adopted in the analysis, our study here is not entirely self-consistent, since the boost factor parametrisation adopted (see Appendix \ref{app:halo.ann}) were determined for a specific minimum subhalo mass.

\section{Comparison against the 8-year dataset}\label{app:8vs12}
  \begin{figure*}
      \centering
      \includegraphics[width=0.49\textwidth]{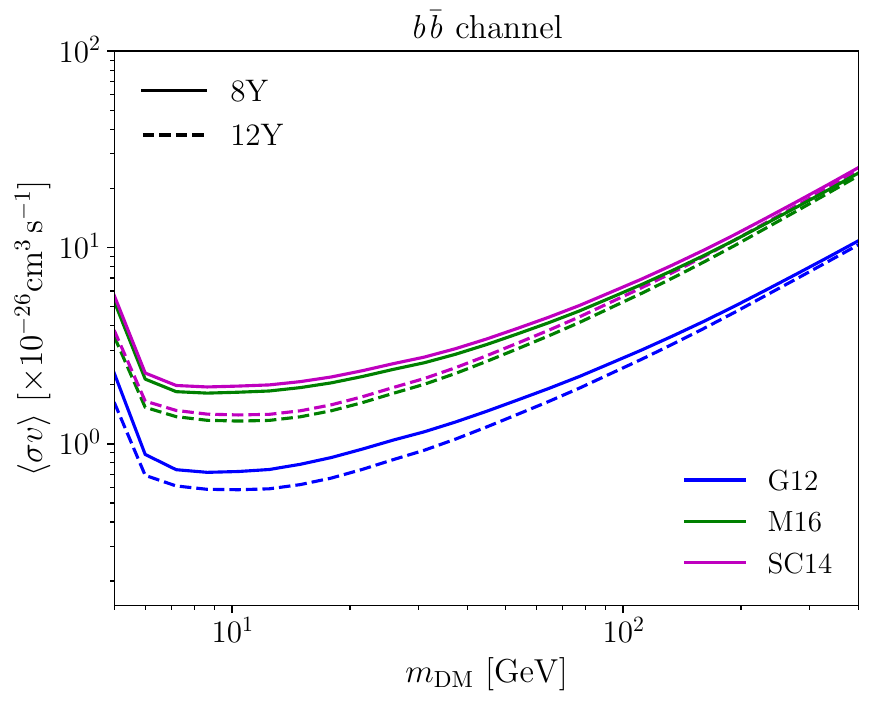}
      \includegraphics[width=0.49\textwidth]{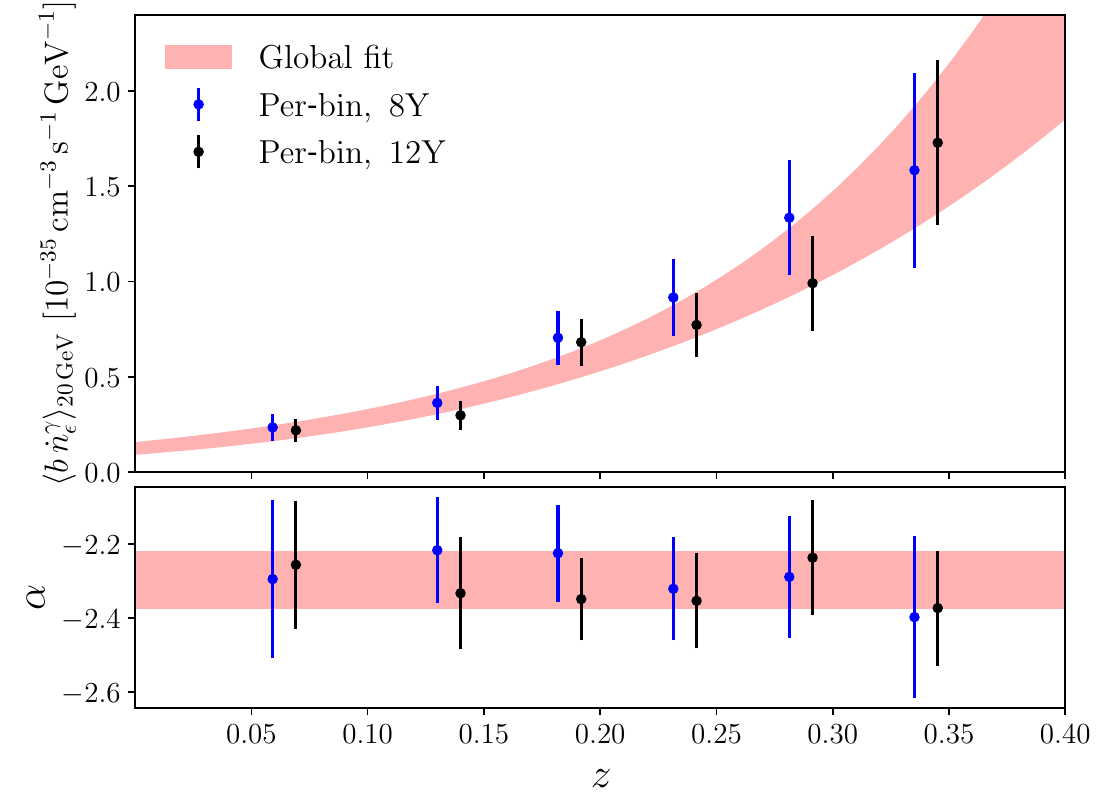}
      \caption{{\sl Left panel:} 95\% upper bounds on the DM annihilation cross section for the $b\bar{b}$ channel, comparing the results found with the 8-year and 12-year datasets in the three substructure regimes. {\sl Right panel:} Constraints on the amplitude and spectral index of the astrophysical model for the \gammaray emissivity (see Eq. \ref{eq:plaw_e_astro}), showing the per-bin constraints found with the 8-year (blue) and 12-year (black) \Fermi datasets.}
      \label{fig:8vs12}
  \end{figure*}
  This Appendix compares the results obtained in this work, using the 12-year \Fermi data, with those obtained using the 8-year dataset in cross-correlation with the same galaxy samples. The overall signal-to-noise ratio of the full set of cross-correlations is reduced by $\sim1$ (e.g. ${\rm SNR}_{\rm astro,Y8}=9.1$, instead of $9.8$), as do, on average, each of the individual cross-correlations presented in Fig. \ref{fig:cls}. We do find larger variations in bins 2, 3 and 6, while bins 1, 4, and 5 remain at roughly the same SNR.
  
  This increase in sensitivity is accompanied by tighter constraints on the dark matter and astrophysical parameters of the models used in this work. Fig. \ref{fig:8vs12} compares the constraints from the 8-year and 12-year datasets on the DM annihilation cross-section (left panel), and on the amplitude and spectral tilt of our astrophysical model (right panel). The upper bound on $\sigv$ improves by a factor $\sim1.4$ with the newer data. The astrophysical constraints are tightened at a similar level, and are compatible with the results obtained with the 12-year data.

\end{document}